\documentclass[%
aip,
amsmath,amssymb,
reprint,%
]{revtex4-1}

\usepackage{graphicx}
\usepackage{dcolumn}
\usepackage{bm}

\usepackage[utf8]{inputenc}
\usepackage[T1]{fontenc}
\usepackage{mathptmx}
\begin{document}

\preprint{AIP/123-QED}

\title[]{Anomalous Hall effect and negative longitudinal magnetoresistance in half-Heusler topological semimetal candidates TbPtBi and HoPtBi}

\author{O. Pavlosiuk}
  \email{o.pavlosiuk@intibs.pl.}
 \affiliation{Institute of Low Temperature and Structure Research, Polish Academy of Sciences, 50-422 Wroc{\l}aw, Poland}
\author{P. Fa{\l}at}%
\affiliation{Institute of Low Temperature and Structure Research, Polish Academy of Sciences, 50-422 Wroc{\l}aw, Poland
}%

\author{D. Kaczorowski}
\affiliation{Institute of Low Temperature and Structure Research, Polish Academy of Sciences, 50-422 Wroc{\l}aw, Poland
}%
\author{P. Wi\'{s}niewski}
\affiliation{Institute of Low Temperature and Structure Research, Polish Academy of Sciences, 50-422 Wroc{\l}aw, Poland
}%

\begin{abstract}
Half-Heusler compounds have attracted significant attention because of their topologically non-trivial electronic structure, which leads to unusual electron transport properties.
We thoroughly investigated the magnetotransport properties of high-quality single crystals of two half-Heusler phases, TbPtBi and HoPtBi, in pursuit of the characteristic features of topologically non-trivial electronic states.
Both studied compounds are characterized by the giant values of transverse magnetoresistance with no sign of saturation in magnetic field up to 14\,T. 
HoPtBi demonstrates the Shubnikov-de Haas effect with two principal frequencies, indicating a complex Fermi surface; the extracted values of carrier effective masses are rather small, $0.18\,m_e$ and $0.27\,m_e$. 
The investigated compounds exhibit negative longitudinal magnetoresistance and anomalous Hall effect, which likely arise from a nonzero Berry curvature. 
Both compounds show strongly anisotropic magnetoresistance, that in HoPtBi exhibits a butterfly-like behavior. 
\end{abstract}

\maketitle

\section{Introduction}

Topological materials, i.e., materials with non-trivial topology of the electronic structure, were very intensively studied during the last decade.~\cite{Armitage2018,Hasan2011} 
Despite this fact, searching for new representatives of this family of compounds is still of great importance because it can provide a route to better understanding of their fascinating physical properties and expand a basis for potential applications. 
Half-Heusler phases crystallizing in the cubic MgAgAs-type crystal structure are renowned for the diversity of their chemical compositions; consequently, they show a very wide gamut of physical properties.~\cite{Graf2011a} 
We focus our investigations mainly on the rare-earth (RE) based half-Heusler phases whose physical properties are very diverse, including antiferromagnetism, superconductivity, heavy-fermion behavior, good thermoelectric performance, and non-trivial topological properties.~\cite{Nakajima2015d,Mun2013,Liu2016a,Sekimoto2006,Pavlosiuk2015,Pavlosiuk2016a,Pavlosiuk2016b,Synoradzki2018,Synoradzki2018a,Wolanska2019,Synoradzki2019}
Particular attention is paid to platinum-bearing half-Heusler
materials, which were found to exhibit the distinctive features of topological semimetals.~\cite{Chadov2010a,Al-Sawai2010,Liu2016a,Hirschberger2016a}
Besides, some of these compounds were reported to superconduct at low temperatures despite very low carrier concentrations.~\cite{Butch2011a,Tafti2013,Pavlosiuk2016b,Pavlosiuk2020_ScPtBi} 
This is especially interesting because the mechanisms of superconductivity in these systems can be unconventional.~\cite{Meinert2015,Kim2018a}
The results of both, electronic structure calculations and angle-resolved photoemission spectroscopy (ARPES) studies, showed that the well-known heavy-fermion system YbPtBi~\cite{Fisk1991} holds triply degenerate fermion points.~\cite{Guo2018}
Moreover, YbPtBi demonstrates the anomalous Hall effect (AHE), arising from the Berry curvature produced by Weyl points, and negative longitudinal magnetoresistance (LMR).~\cite{Guo2018} 
The latter is a smoking gun signature of chiral magnetic anomaly (CMA), which should take place in topological semimetals due to charge pumping between two Weyl nodes with opposite chiralities.~\cite{Son2013}
The occurrence of negative LMR induces the substantial value of anisotropic magnetoresistance (AMR), a physical property that can find its application in magnetic field sensing technology.~\cite{Grosz2017}     
For YPtBi and LuPtBi, the existence of states with linear energy dispersion, typical for topological materials, was confirmed directly by ARPES.~\cite{Liu2016a,Hosen2018}    
Recently, another topologically non-trivial half-Heusler compound, viz., GdPtBi, stirred the most interest as the first example of a magnetic-field-induced Weyl semimetal.~\cite{Hirschberger2016a}
Its topologically non-trivial character was proven by the observation of negative LMR,~\cite{Hirschberger2016a} large AHE,~\cite{Suzuki2016} and electronic structure calculations.~\cite{Hirschberger2016a}
Interestingly, in Ref.~\onlinecite{Hirschberger2016a}, the authors concluded that zero-gap semiconductors with strong spin-orbit coupling are appropriate candidates for the magnetic-field-induced Weyl semimetals. 
This prediction motivated us to undertake our own investigations of other half-Heusler RE-based platinum bismuthides.
    
Here, we report the results of our studies on TbPtBi and HoPtBi with antiferromagnetic ground states.
To date, both compounds were characterized for their crystal structure,~\cite{Haase2002} magnetic behavior, and basic electrical properties (resistance, magnetoresistance and Hall effect).~\cite{Canfield1991,Chen2020,Singha2019a,Zhu2020,Chen2020a}
TbPtBi was recently reported to demonstrate negative LMR and anomalous Hall angle (AHA), whereas only the former phenomenon was found in HoPtBi.~\cite{Singha2019a,Zhu2020} 
Our thorough investigations of the electrical transport and thermodynamic properties (magnetization and specific heat) of both materials were carried out on high-quality single crystals.
The observation of AHE and negative LMR strongly supports the existence of topologically non-trivial states in the two phases. 
Besides, their AMR retains significant value in wide ranges of temperature and magnetic field. 

\section{Experimental methods}

Single crystals of TbPtBi and HoPtBi were grown from Bi-flux as reported in Ref.~\onlinecite{Pavlosiuk2016b}. 
They had metallic luster, were stable in air, and their dimensions were up to $1.5\!\times\!1.5\!\times\!1.5$\,mm$^3$. 
Their chemical composition was investigated on a FEI scanning electron microscope equipped with an EDAX Genesis XM4 spectrometer. 
The obtained chemical compositions Tb$_{33.78}$Pt$_{34.92}$Bi$_{31.30}$ and Ho$_{33.12}$Pt$_{33.58}$Bi$_{33.12}$ were very close to nominal ones. 
X-ray diffraction experiments, performed on powdered single crystals, confirmed the MgAgAs-type crystal structure with the lattice parameters $a=6.665$\,\AA\, and $a=6.632$\,\AA\, for TbPtBi and HoPtBi, respectively, in good agreement with the literature data.~\cite{Haase2002} 
Single crystals' quality and their orientation were determined by the Laue backscattering technique using a Proto LAUE-COS system (exemplary Laue patterns are shown in the supplementary material).
Magnetization and magnetic susceptibility measurements were carried out employing a Quantum Design MPMS-XL magnetometer equipped with a $^3$He refrigerator. 
Electrical resistivity, magnetoresistance, anisotropic magnetoresistance, Hall effect and heat capacity were investigated using a Quantum Design PPMS platform. 
Electrical contacts for the transport measurements were prepared with 50\,$\mu$m-diameter silver wires attached to the samples with silver paste.

\section{Results and Discussion}

\subsection{\label{sec:level2}Magnetic properties and specific heat}

\begin{figure}[h]
	\includegraphics[width=0.46\textwidth]{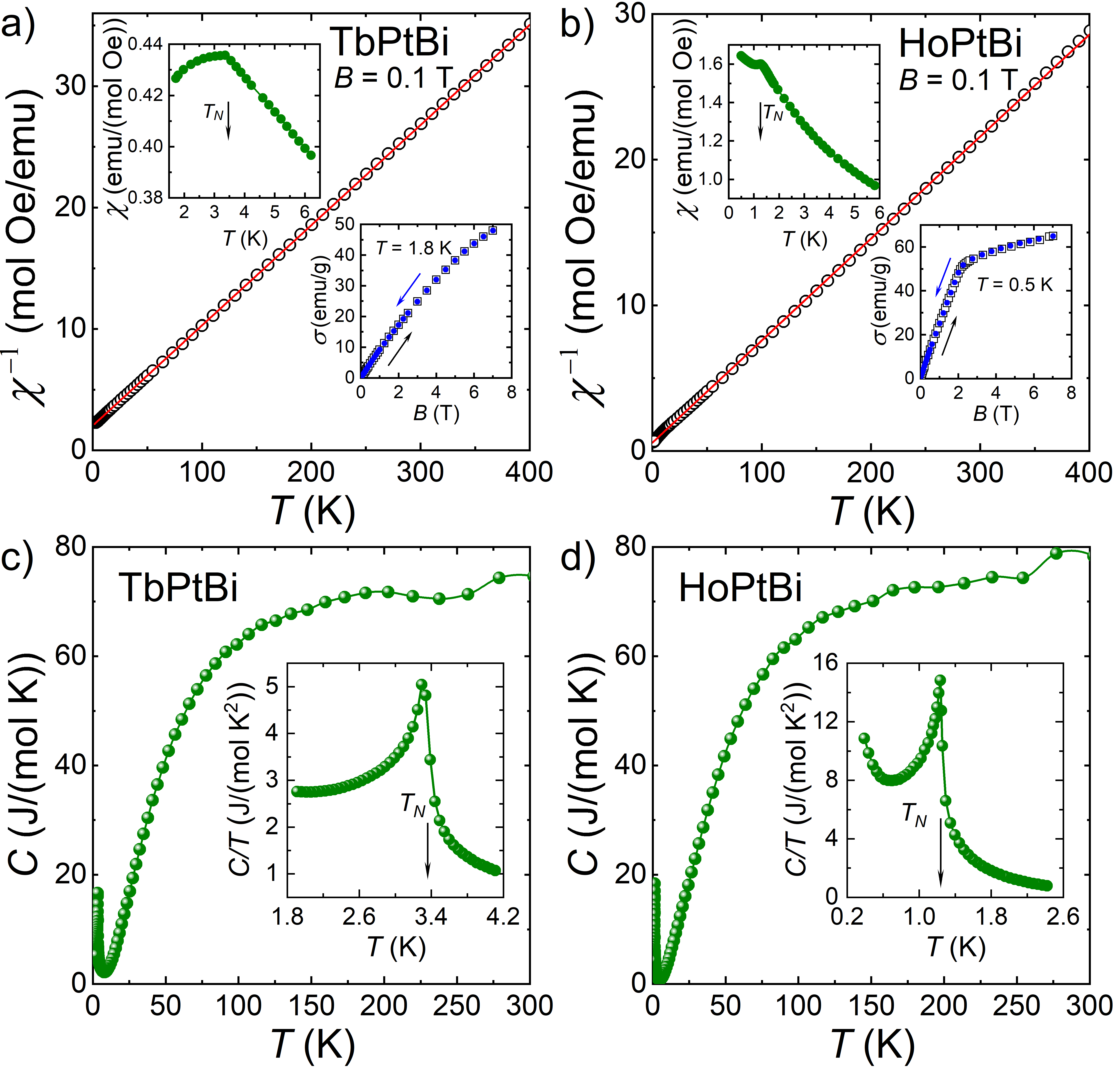}
	\caption{Temperature dependence of the reciprocal magnetic susceptibility of TbPtBi (a) and HoPtBi (b). Red solid lines represent Curie-Weiss law fits (Eq.~\ref{Eq_Curie_Weiss}). Upper insets: low-temperature magnetic susceptibility measured in magnetic field 0.1\,T. Lower insets: the magnetic field variation of the magnetization recorded at 1.8\,K for TbPtBi and at 0.5\,K for HoPtBi, with increasing (open black squares) and decreasing (full blue circles) magnetic field. The temperature dependence of the specific heat of TbPtBi (c) and HoPtBi (d). Insets: low-temperature dependences of specific heat over the temperature, black arrows mark N\'{e}el temperatures. 
		\label{Fig1}}
\end{figure}

According to the previous report, TbPtBi orders antiferromagnetically below the N\'{e}el temperature ($T_N$) of 3.4\,K, while for HoPtBi $T_N$ smaller than 1.2\,K was anticipated.~\cite{Canfield1991}
The results of our magnetic measurements confirmed the antiferromagnetic ordering in both compounds below $T_N$ equal to 3.36\,K and 1.26\,K for TbPtBi and HoPtBi, respectively (see the upper insets to Fig.~\ref{Fig1}a,b). 
In the low-temperature range, the temperature dependence of the magnetic susceptibility, $\chi$, of HoPtBi exhibits a peak at $T_N$, but it is followed by increasing $\chi$ with further lowering $T$, similar to $\chi(T)$ of GdPtBi.~\cite{Canfield1991}
This anomalous behavior can be attributed to some reorientation of magnetic moments.~\cite{Behr2001}

Above about 25\,K, the inverted magnetic susceptibility of both compounds linearly increases with $T$, obeying the Curie-Weiss law:
\begin{equation}
\chi^{-1}(T)=\frac{3k_B(T-\theta_P)}{N_A\mu^2_{e\!f\!f}},
\label{Eq_Curie_Weiss}
\end{equation}
where $k_B$ is the Boltzmann constant, $N_A$ is the Avogadro constant, $\theta_P$ is the paramagnetic Curie temperature and $\mu_{e\!f\!f}$ is the effective magnetic moment. 
The results of fitting Eq.~\ref{Eq_Curie_Weiss} to the experimental data are shown as red solid lines in the main panels of Fig.~\ref{Fig1}a,b, and the parameters $\mu_{e\!f\!f}$ and $\theta_p$ are listed in Table~\ref{Tab_Curie_Weiss_param}; their values are close to those reported in the literature.~\cite{Canfield1991,Singha2019a}
The obtained values of $\mu_{e\!f\!f}$ point to the trivalent character of Tb and Ho ions.  
\begin{table}[h]
	\centering
	\caption{N\'{e}el temperatures, $T_N$; theoretical effective magnetic moments, $\mu^{theor}_{e\!f\!f}$; parameters obtained from the Curie-Weiss law fit: effective magnetic moments, $\mu_{e\!f\!f}$, paramagnetic Curie temperatures, $\theta_P$ for TbPtBi and HoPtBi}
	\begin{tabular*}{0.45\textwidth}{@{\extracolsep{\fill}}*{5}{c}} \hline\hline
		Compound & $T_N$ (K)& $\mu^{theor}_{e\!f\!f}$ ($\mu_B$) & $\mu_{e\!f\!f}$ ($\mu_B$) & $\theta_P$ (K) \\\hline
		TbPtBi & 3.36 & 9.72 & 9.85 & -24.6 \\
		HoPtBi & 1.26 & 10.6 & 10.7 & -8.2 \\
		\hline\hline
	\end{tabular*}\label{LMtable}
	\label{Tab_Curie_Weiss_param}
\end{table}

The lower insets of Fig.~\ref{Fig1}a,b show the magnetic field dependent magnetization, $\sigma(B)$, measured at temperatures below $T_N$. 
The curves obtained with increasing (black open squares) and decreasing (blue filled circles) magnetic fields are identical, therefore there is no hysteresis.

The results of specific heat, $C$, measurements, shown in Fig.~\ref{Fig1}c,d, confirmed the antiferromagnetic order in both compounds. 
Clear $\lambda$-shaped peaks, in $C(T)/T$ were found at 3.29\,K and 1.24\,K for TbPtBi and HoPtBi, respectively (see the insets of Fig.~\ref{Fig1}c,d), in agreement with $T_N$ obtained from the magnetic susceptibility data. 
$C(T)/T$ of HoPtBi exhibits an upturn below $T=0.7$\,K, which can be attributed to the nuclear Schottky effect. 
Similar behavior, arising from a large nuclear quadrupole moment of Ho, was reported for pure holmium,~\cite{Holmstrom1969} as well as for several Ho-bearing compounds, e.g., borocarbides\cite{Rapp1999} or the half-Heusler representative HoPdBi.~\cite{Nakajima2015d,Pavlosiuk2016a} 

\subsection{\label{sec:level2}Electrical resistivity}

\begin{figure}
	\includegraphics[width=0.46\textwidth]{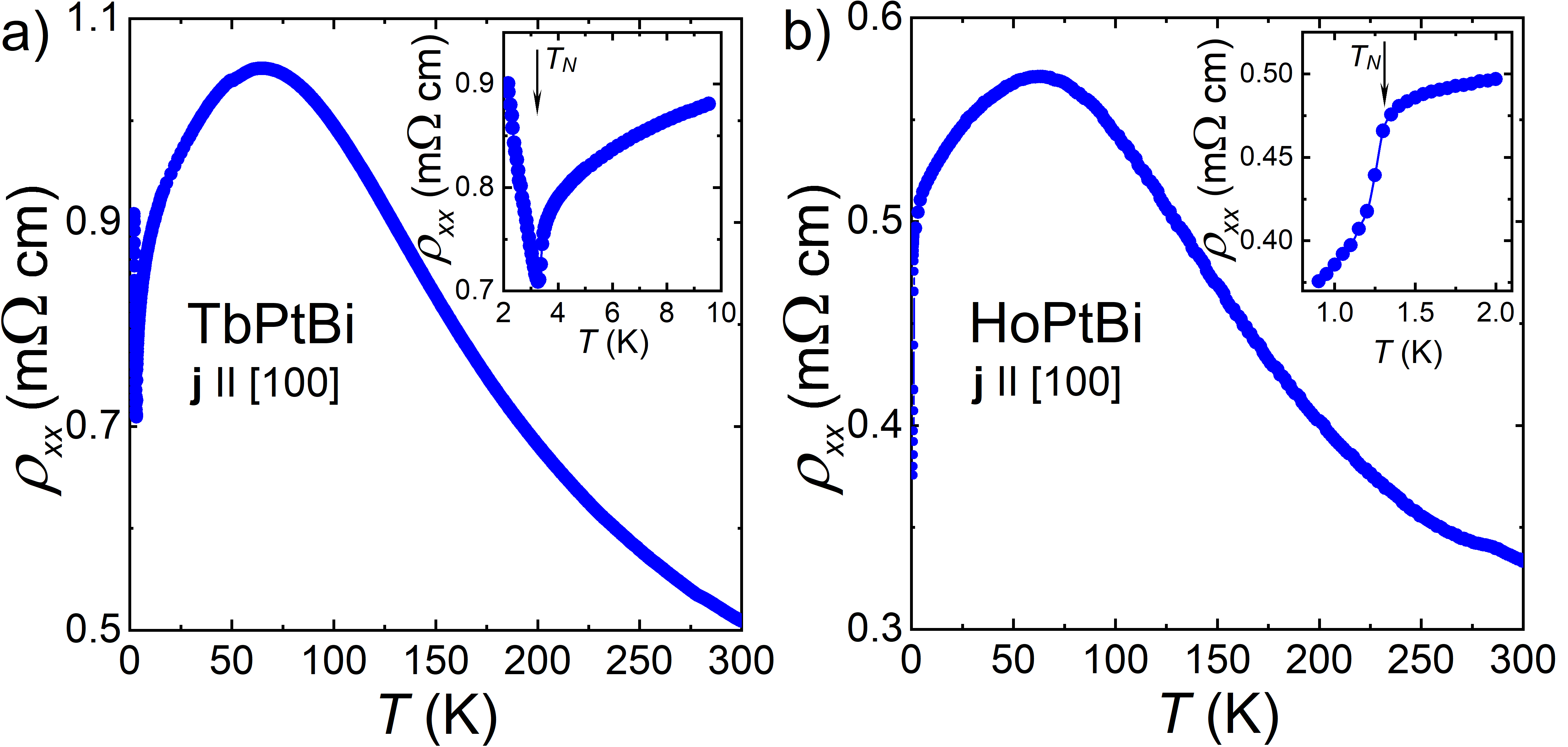}
	\caption{Temperature dependence of the electrical resistivity of TbPtBi (a) and HoPtBi (b), measured along the [100] crystallographic direction. Insets show the low-temperature data; arrows indicate the antiferromagnetic phase transitions.
		\label{Fig2}}
\end{figure}

The temperature dependence of the electrical resistivity, $\rho_{xx}$, of TbPtBi and HoPtBi is presented in Fig.~\ref{Fig2}a,b. 
For both compounds, $\rho_{xx}(T)$ is similar to those observed in other half-Heusler bismuthides.~\cite{Pavlosiuk2015,Pavlosiuk2016a,Pavlosiuk2016c,Nakajima2015d,Gofryk2011,Pan2013a,Mun2016a} 
Initially, $\rho_{xx}$  increases with decreasing temperature, achieving the maximum values of $0.57$\,\rm{m}$\Omega$\,\rm{cm} at $T^*=62$\,K and $1.05$\,\rm{m}$\Omega$\,\rm{cm} at $T^*=65$\,K for TbPtBi and HoPtBi, respectively. 
Below $T^*$, $\rho_{xx}$ decreases with decreasing temperature; for HoPtBi, down to the lowest temperatures studied and for TbPtBi, down to $T_N$ only. 
The antiferromagnetic phase transition in HoPtBi manifests itself as a distinct kink in $\rho_{xx}(T)$ (see the inset to Fig.~\ref{Fig2}b). 
In contrast, as shown in Fig.~\ref{Fig2}a, $\rho_{xx}(T)$ of TbPtBi shows a sharp dip at $T_N$; such anomalies can be attributed to the reconstruction of the Fermi surface at transition to the antiferromagnetic state. 
Similar behavior was observed for other half-Heusler compounds, e.g. GdPtBi and DyPtBi.~\cite{Mun2016a,Canfield1991}
In the paramagnetic region, the overall $\rho_{xx}(T)$ behavior of both compounds resembles that typical for semimetals or narrow-gap semiconductors.~\cite{Dornhaus1983} 
In particular, the results obtained for HoPtBi are fairly similar to the literature data.~\cite{Canfield1991,Chen2020}
However, $\rho_{xx}(T)$ of TbPtBi notably differs from the behavior reported in Refs.~\onlinecite{Zhu2020, Singha2019a}, with regard to both the resistivity magnitude and the shape of the $\rho_{xx}(T)$ variation. 
The differences can be attributed to various quality of the investigated single crystals, as was established before, e.g., for YPtBi,~\cite{Butch2011a,Bay2012a,Bay2014,Pavlosiuk2016b} GdPtBi~\cite{Hirschberger2016a} and LuPdBi.~\cite{Pavlosiuk2015,Xu2014}
Most recently, we studied in detail the electronic transport in the half-Heusler phase ScPtBi, where even small amount of vacancies or antisite defects can drastically change the electronic band structure and, as a consequence, its transport properties.~\cite{Pavlosiuk2020_ScPtBi}

\subsection{\label{sec:level2}Magnetoresistance}

\begin{figure}[b]
	\includegraphics[width=0.46\textwidth]{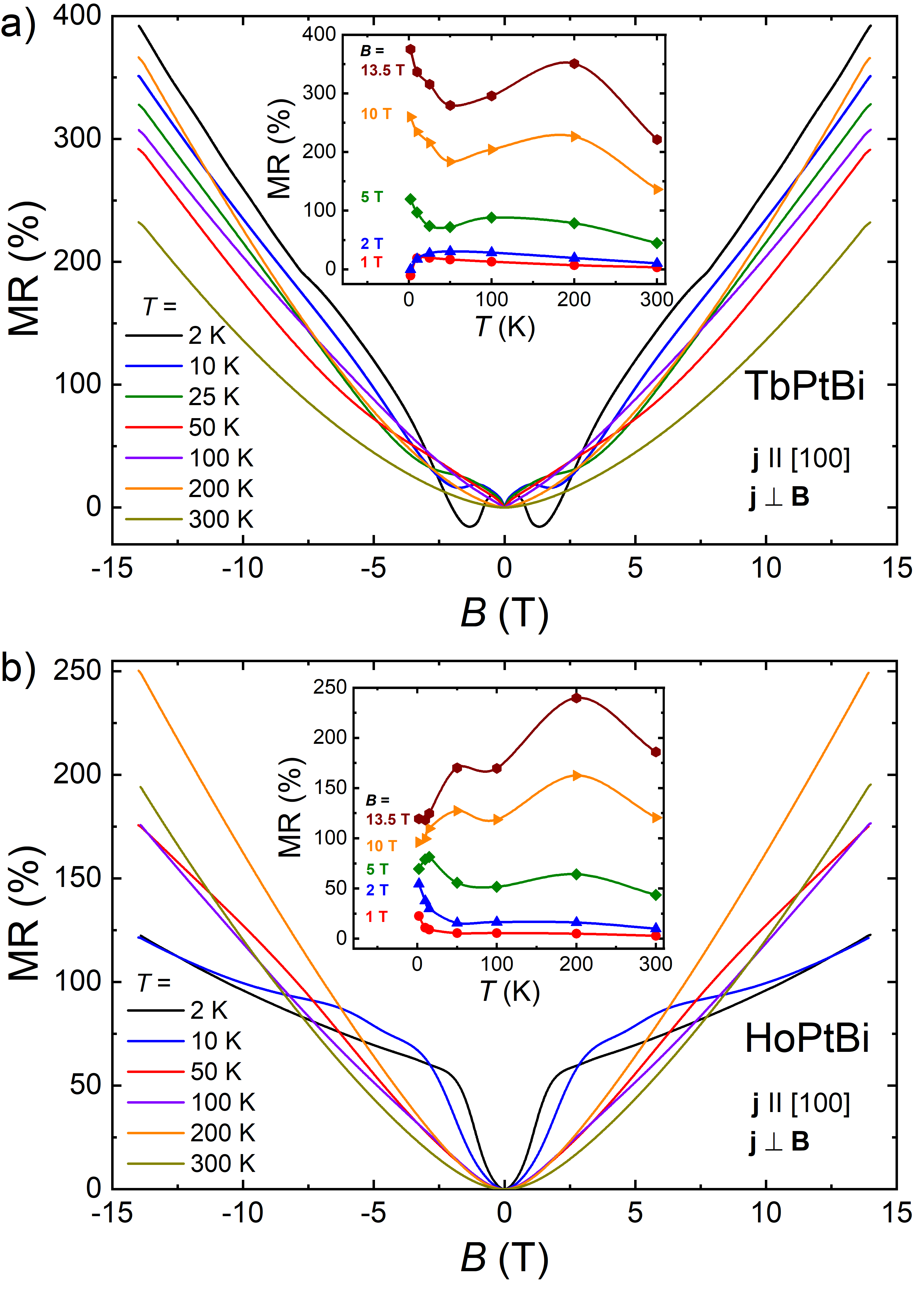}
	\caption{Transverse magnetoresistance as a function of magnetic field for TbPtBi (a) and HoPtBi (b), recorded at several temperatures. Insets show the temperature dependence of the transverse magnetoresistance measured in several constant values of magnetic field. Solid lines are guides for the eye.
		\label{Fig3}}
\end{figure}

Magnetoresistance, MR\,$=\rho_{xx}(B)/\rho_{xx}(0)-1$, isotherms measured in the transverse geometry ($\rm\bf{j\perp B}$) are shown in Fig.~\ref{Fig3}a and Fig.~\ref{Fig3}b, for TbPtBi and HoPtBi, respectively. 
Both compounds show positive and non-saturating magnetoresistance of large magnitude, most probably due to the existence of charge carriers originating from different electronic bands.~\cite{Pippard2009}
Charge carriers of different types have previously been found in other RE-based half-Heusler compounds, such as YPtBi, LuPtBi and GdPtBi.~\cite{Xu2017,Meinert2015,Schindler2018b}
MR of HoPtBi achieves its maximum value of $250\%$ at 200\,K in $B=14$\,T. 
For TbPtBi, the maximum value of MR equals $392\%$ and it was found at $T=2$\,K and in $B=14$\,T, but if we consider the paramagnetic region only, the largest value ($\rm{Fig. 3}, 366\%$) is observed at $T=200$\,K, alike for HoPtBi. 
It is worth noting that MR found here for TbPtBi exceeds by an order of magnitude MR reported for this compound in Refs.~\onlinecite{Singha2019a,Chen2020}. 
In addition, MR of HoPtBi is larger than that reported in Ref.~\onlinecite{Chen2020}, but the difference is much smaller.
At low temperatures and in a weak-magnetic-field region, MR($B$) of HoPtBi forms a hump, whereas MR($B$) of TbPtBi makes a dip with negative MR at $T = 2$\,K. 
Similar features were established for these compounds in Refs.~\onlinecite{Chen2020,Singha2019a,Zhu2020,Chen2020a}. 
Most remarkably, a comparable dip and negative values of MR were observed also for GdPtBi,~\cite{Hirschberger2016a,Shekhar2018,Schindler2018b} and attributed either to the reconstruction of the Fermi surface (due to an external magnetic field~\cite{Hirschberger2016a} or due to an exchange magnetic field~\cite{Shekhar2018}) or to the magnetic interactions of itinerant charge carriers with localized Gd $4\!f$ magnetic moments and paramagnetic impurities.~\cite{Schindler2018b} 

MR of the studied compounds changes with varying temperature in a non-monotonous manner (see insets to Fig.~\ref{Fig3}a,b), alike for other RE-based half-Heusler compounds.~\cite{Mun2016a,Chen2020,Chen2020a}
Such a behavior may suggest that the carrier concentration and carrier mobility are strongly dependent on the magnetic field and temperature. 
In concert with this presumption, the transverse magnetoresistance of TbPtBi and HoPtBi was found to disobey the Kohler's scaling (see Fig.~S2 of the supplementary material).

\subsection{\label{sec:level2}Shubnikov-de Haas oscillations}

\begin{figure}[t]
	\includegraphics[width=0.46\textwidth]{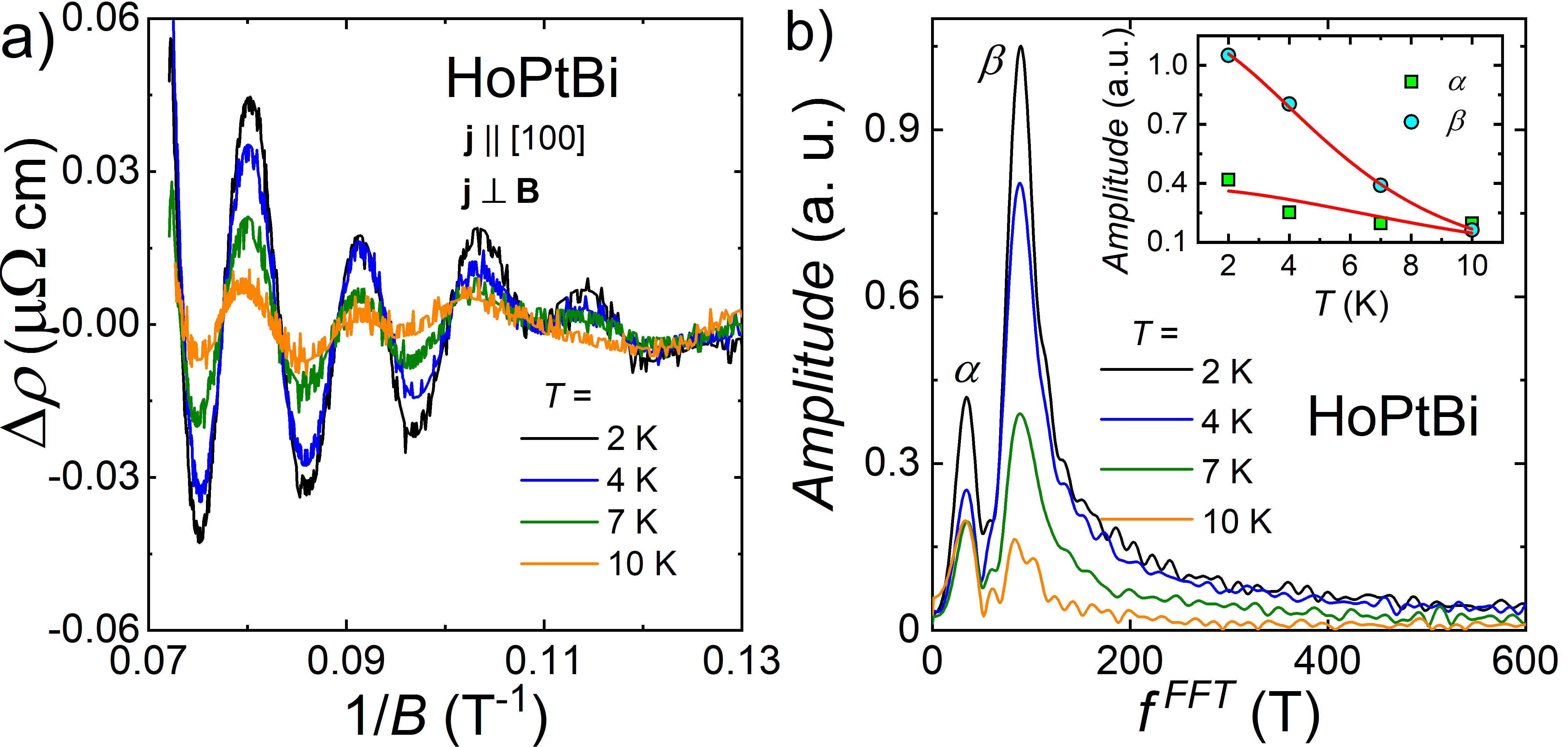}
	\caption{(a) Oscillating part of the electrical resistivity measured in transverse magnetic field at several different temperatures. (b) The fast Fourier transform analysis of the data presented in (a). Inset: temperature dependences of the amplitudes of two principal frequencies in the FFT spectra. Red solid lines represent the fits of Eq.~\ref{LK_eq} to the experimental data.
		\label{Fig4}}
\end{figure}

Upon subtracting a smooth (second order polynomial) background from the $\rho(B)$ of HoPtBi, the pronounced Shubnikov-de Haas (SdH) oscillations were identified (see Fig.~\ref{Fig4}a). 
The fast Fourier transform (FFT) analysis yielded spectra with two distinct peaks at $F_{\alpha}=34.48$\,T and $F_{\beta}=89.44$\,T (see Fig.~\ref{Fig4}b).
The observation of two principal frequencies contrasts with that reported in Ref.~\onlinecite{Chen2020a}, where only one principal frequency of 40\,T was noticed. 
The observation of two frequencies may stand for the complexity of the Fermi surface in HoPtBi or for the multi-carrier type of conductivity, where at least two different bands contribute to the electronic transport properties. 
According to the recent results of electronic structure calculations for HoPtBi, its Fermi level crosses both, electron- and hole-like bands,~\cite{Chen2020a} similar to other RE-based platinum bismuthides.~\cite{Shekhar2018,Schindler2018b,Xu2017} 
Assuming roughly that the Fermi pockets in HoPtBi are spherical, the Fermi wave vectors, $k_{F,i}$, were calculated using the Onsager's relation: $F_i=\hbar S/2\pi e$, where $S=\pi k^2_{F,i}$ is the area of the $i$-th Fermi pocket cross-section, and $\hbar$ is the reduced Planck constant ($i=\alpha, \beta$). 
From the so-obtained values $k_{F,\alpha}=0.032$\,\AA$^{-1}$ and $k_{F,\beta}=0.052$\,\AA$^{-1}$, the carrier concentrations $n^{\rm{SdH}}_{\alpha}=1.15\times10^{18}\,\rm{cm}^{-3}$ and $n^{\rm{SdH}}_{\beta}=4.79\times10^{18}\,\rm{cm}^{-3}$ were computed using the relation $n^{\rm{SdH}}_{i}=V_{F,i}/(4\pi^3)$, where $V_{F,i}$ stands for the Fermi pocket volume. 
It is worth noting that both values of $n^{\rm{SdH}}_{i}$ are very close to those obtained from the analysis of the Hall effect data measured at $T=10$\,K (see below). 
We also calculated the effective masses of carriers in the two Fermi pockets by fitting the standard equation that describes the thermal damping of SdH oscillations (Eq.~\ref{LK_eq}) to the temperature dependences of the FFT amplitudes (see inset to Fig.~\ref{Fig4}b).
\begin{equation}
R_i(T)=(\lambda m^*_iT/B_{\rm{eff}})/\sinh(\lambda m^*_iT/B_{\rm{eff}}), (i=\alpha, \beta)
\label{LK_eq}
\end{equation}
In Eq.~\ref{LK_eq}, $m^*_i$ stands for the effective cyclotron mass and $B_{\rm{eff}}=2/(1/B_1 + 1/B_2)=9.93$\,T. 
This value of $B_{\rm{eff}}$ was obtained for $B_1=14$\,T and $B_2=7.7$\,T, the limits of the magnetic field range in which our FFT analysis was carried out.
The so-obtained effective masses are $m^*_{\alpha}=0.18\,m_e$ and $m^*_{\beta}=0.27\,m_e$, similar to those reported for other half-Heusler phases.~\cite{Pavlosiuk2016b,Pavlosiuk2016a,Kozlova2005,Butch2011a} 
For the other half-Heusler bismuthide studied, i.e., TbPtBi, no clear SdH oscillations were detected in the magnetotransport data, most likely because of the insufficient quality of the single crystal investigated. 

\subsection{\label{sec:level2}Hall effect}

\begin{figure}[t]
	\includegraphics[width=0.46\textwidth]{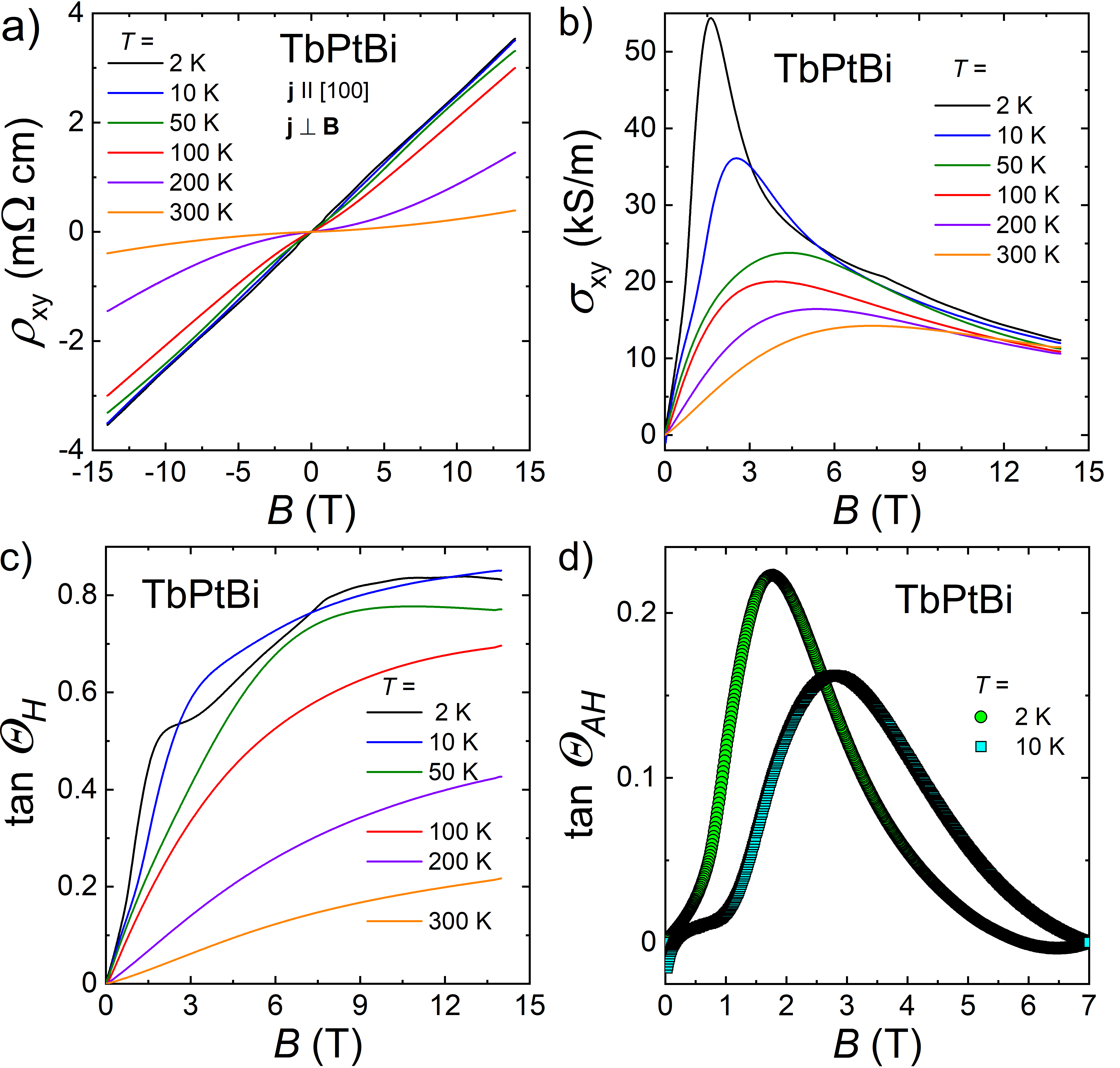}
	\caption{Hall resistivity (a) and the Hall conductivity (b) of TbPtBi as a function of magnetic field at several different temperatures. (c) The tangent of the Hall angle as a function of magnetic field. (d) The tangent of the anomalous Hall angle as a function of magnetic field at 2\,K and 10\,K.
		\label{Fig5}}
\end{figure}

\begin{figure}[h]
	\includegraphics[width=0.46\textwidth]{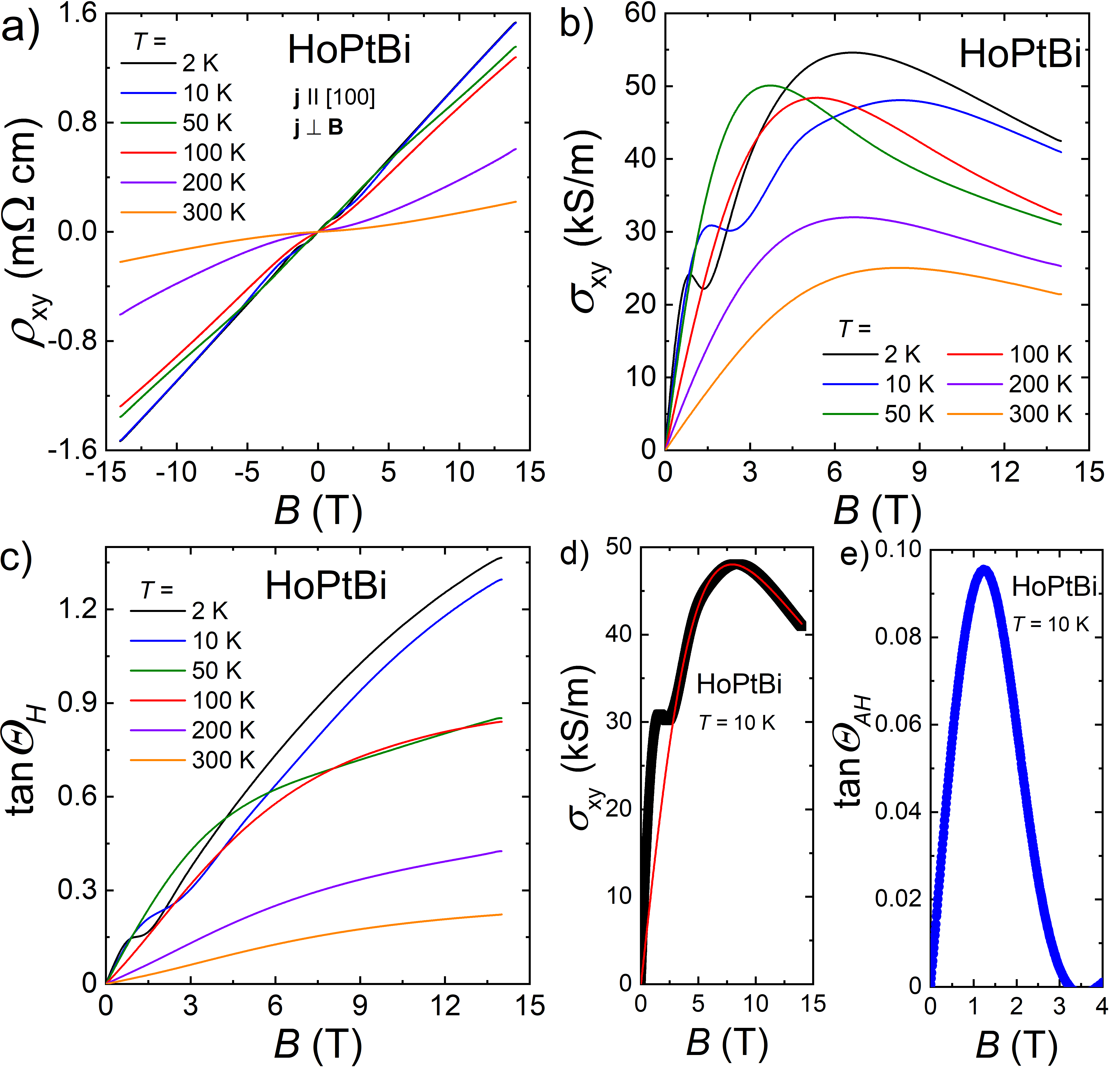}
	\caption{Hall resistivity (a) and the Hall conductivity (b) of HoPtBi as a function of magnetic field at several different temperatures. (c) The tangent of the Hall angle as a function of magnetic field. (d) The Hall conductivity as a function of magnetic field at 10\,K; the red solid line is the generated theoretical curve that corresponds to the single-band Drude model (for details, see text). (e) The tangent of anomalous Hall angle as a function of magnetic field at 10\,K. 
		\label{Fig6}}
\end{figure}

Figures~\ref{Fig5}a and \ref{Fig6}a present the Hall resistivity, $\rho_{xy}$, as a function of magnetic field measured for TbPtBi and HoPtBi, respectively, at several constant temperatures. 
$\rho_{xy}$ of TbPtBi is an order of magnitude larger than that reported in Refs.~\onlinecite{Singha2019a,Zhu2020}, suggesting that our sample had a smaller carrier concentration.
A similar magnitude of $\rho_{xy}$ was found for HoPtBi.    
In both compounds, hole-type carriers dominate the transport properties. 
The isotherms $\rho_{xy}(B)$ are curvilinear, which may arise from the strong dependence of the electronic structure on the magnetic field or point to the multi-band type of conductivity, typical for half-Heusler compounds.~\cite{Pavlosiuk2019a,Pavlosiuk2016b} 
Remarkably, non-linear $\rho_{xy}(B)$ is also a strong indication of AHE.~\cite{Zhu2020,Suzuki2016,Shekhar2018} 

In an attempt to determine the origin of the Hall effect in the studied compounds, we used the multi-band Drude model:
\begin{equation}
\sigma_{xy}=\sum_{i=1}^pe\,n_i\mu_i^2B/[1+(\mu_iB)^2],
\label{Eq_two-band_Drude}
\end{equation} 
where $p$ is the total number of Fermi pockets, $e$ is the elementary charge, $n_i$ and $\mu_i$ are the carrier concentration and mobility of the $i$-th band, respectively.
First, we converted $\rho_{xy}$ into the Hall conductivity, $\sigma_{xy}$, using the relation: $\sigma_{xy}=\rho_{xy}/(\rho_{xy}^2+\rho_{xx}^2)$, (see Fig.~\ref{Fig5}b and Fig.~\ref{Fig6}b). 
Distinct peaks in the isotherms measured at $T=2$\,K and $T=10$\,K occur in $B=0.87$\,T and $B=1.65$\,T for HoPtBi, and in $B=1.62$\,T and $B=2.51$\,T for TbPtBi.
Next, we fitted the $\sigma_{xy}(B)$ data with Eq.~\ref{Eq_two-band_Drude} using $p$ values equal to 1 and 2, which correspond to the single- and two-band Drude models, respectively.
We found that none of these models approximates the experimental data satisfactorily at low temperatures. 
A reasonably good description of $\sigma_{xy}(B)$ was obtained only for $T\geq$\,50\,K and $T\geq$\,100\,K for HoPtBi and TbPtBi, respectively (for details see the supplementary material).
It should be noted that for $T=50\,$K and $T=100$\,K for HoPtBi and TbPtBi, respectively, $\sigma_{xy}(B)$ can be described by single- and two-band Drude models, whereas at higher temperatures, only the single-band model gives satisfactory results of the fitting. 
Parameters obtained from the two-band Drude model fits suggest that both bands are hole-like in both compounds. 
This result is consistent with the electronic band structure of TbPtBi reported in Ref.~ \onlinecite{Zhu2020}.  
At lower temperatures, additional contribution to the total Hall conductivity was noticed, which might be attributed to AHE.~\cite{Suzuki2016,Shekhar2018,Singha2019a} 
To check this conjecture, we calculated the tangent of the Hall angle, $\tan\Theta_{H}=\sigma_{xy}/\sigma_{xx}$, where $\sigma_{xy}$ and $\sigma_{xx}=\rho_{xx}/(\rho_{xy}^2+\rho_{xx}^2)$ are tensor components of the electrical conductivity;~\cite{Hurd1972} 
here, it should be noted that in a few recent reports, related to AHE in half-Heusler compounds,~\cite{Suzuki2016,Shekhar2018,Singha2019a,Guo2018} the ratio $\sigma_{xy}/\sigma_{xx}$ has mistakenly been called the anomalous Hall angle instead of the tangent of the anomalous Hall angle.

The obtained $\tan\Theta_{H}(B)$ dependences are presented in Fig.~\ref{Fig5}c and Fig.~\ref{Fig6}c for TbPtBi and HoPtBi, respectively. 
For both compounds, the magnitude of $\tan\Theta_{H}$ increases with increasing magnetic field and with decreasing temperature. 
The maximum values of $\tan\Theta_{H}$ are $\sim0.8$ and $\sim1.4$ for TbPtBi and HoPtBi, respectively. 
These values are large and similar to those reported for other half-Heusler phases.~\cite{Singha2019a,Suzuki2016,Shekhar2018}
For TbPtBi, the contribution of $\tan\Theta_{AH}$ was extracted by subtracting normalized $\tan\Theta_{H}(B)$, observed at $T=100$\,K in $B=7$\,T, from the data measured below this temperature. 
The resulting curves are shown in Fig.~\ref{Fig5}d. 
The maximum value $\tan\Theta_{AH}=0.23$ was found at $T=2$\,K in $B=1.78$\,T, and it is somewhat smaller than that reported before for TbPtBi.~\cite{Singha2019a,Zhu2020}
$\tan\Theta_{AH}$ decreases with temperature increasing and the maximum position shifts to higher magnetic fields, in agreement with the literature data.~\cite{Suzuki2016,Singha2019a,Zhu2020} 

In the case of HoPtBi, the above method of extracting, which was also frequently used in recent studies,~\cite{Shekhar2018,Singha2019a,Suzuki2016,Zhu2020} appeared ineffective, and hence a different approach was applied. 
First, $\sigma_{xy}(B)$ was fitted with the single-band Drude model in the magnetic field range from 3\,T to 14\,T. 
The two-band Drude model was over-parametrised relative to the data set, and therefore could not be used for the experimental data approximation. 
Next, the obtained fitting parameters $n=4.77\!\times\!10^{18}\,\rm{cm^{-3}}$ and $\mu=1257\,\rm{cm^2V^{-1}s^{-1}}$ were used to generate the theoretical curve for the magnetic field range form zero to 14\,T (red solid line in Fig.~\ref{Fig6}d). 
Subsequently, that function was subtracted from the experimental data and the resulting curve was divided by $\sigma_{xx}(B)$. 
As the result, $\tan\Theta_{AH}$ was derived (see Fig.~\ref{Fig6}e). 
As can be inferred from the figure, $\tan\Theta_{AH}$ in HoPtBi shows a maximum value of $\sim\!0.1$ in $B=1.24$\,T.

The origin of AHE in half-Heusler compounds is still under debate. 
In Ref.~\onlinecite{Suzuki2016}, the authors suggested that both avoided band-crossing and appearance of Weyl nodes induced by the magnetic field can contribute to the Berry curvature, which is the source of AHE in GdPtBi. 
In turn, in Ref.~\onlinecite{Shekhar2018}, the existence of Weyl nodes was claimed as the only origin of AHE in the same material. 
The most recent theory attributed AHE observed in TbPtBi to the Berry curvature generated by anticrossing of spin-split bands near the Fermi level.~\cite{Zhu2020} 
The same authors suggested that this explanation can be suitable for AHE in GdPtBi as well.

\subsection{\label{sec:level2}Angle dependent magnetoresistance}

\begin{figure*}[t]
	\includegraphics[width=1.0\textwidth]{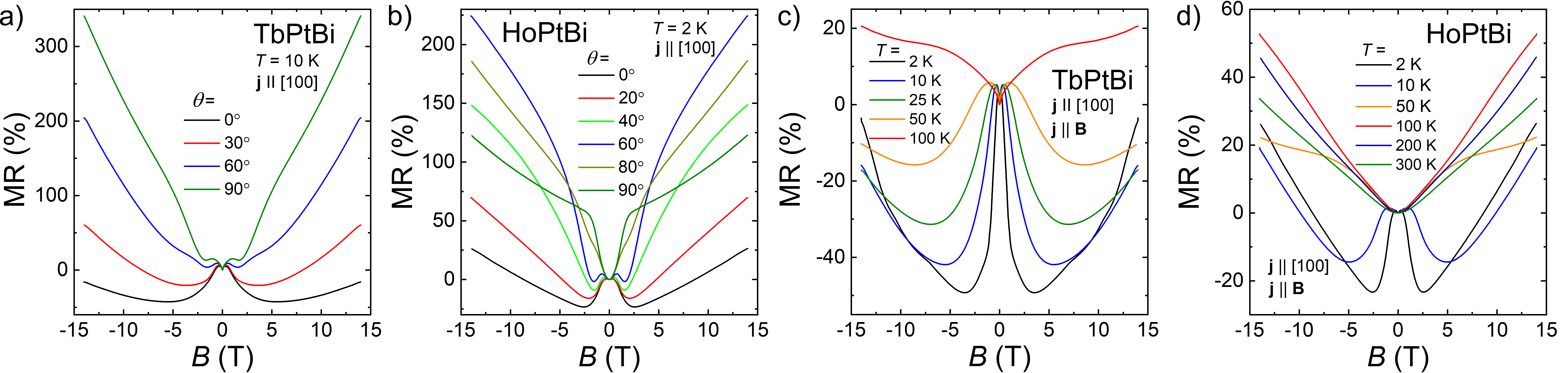}
	\caption{Magnetoresistance vs the magnetic field applied at different angles to the current direction at $T=10$\,K for TbPtBi (a) and at $T=2$\,K for HoPtBi (b). Magnetoresistance vs the magnetic field applied parallel to the current direction at several different temperatures for TbPtBi (c) and HoPtBi (d).
		\label{Fig7}}
\end{figure*}

As suggested in Refs.~\onlinecite{Suzuki2016,Shekhar2018}, the observation of AHE in GdPtBi can be directly related to the existence of Weyl nodes.
Another indication of the presence of Weyl fermions is CMA, which was reported in several other half-Heusler bismuthides, REPtBi (RE = Nd, Gd, Tb, Ho, and Yb).~\cite{Hirschberger2016a,Guo2018,Shekhar2018,Singha2019a,Chen2020a} 
To check if TbPtBi and HoPtBi exhibit CMA, we performed the measurements of MR as a function of magnetic field at constant temperature and at different angles $\theta$ at which the external magnetic field was applied. 
The results of these experiments are shown in Fig.~\ref{Fig7}a,b. 
To eliminate the impact of antiferromagnetic order on the observed properties, the experiments were carried out at temperatures above $T_N$. 
In the weak-magnetic-field region, MR exhibits a small increase with increasing $B$ for all the studied $\theta$ values.
The origin of this behavior can be due to the weak antilocalization (WAL) effect, previously observed in several topological semimetals~\cite{Kim2013a,Lv2017,Huang2015} and many half-Heusler compounds.~\cite{Chen2020,Pavlosiuk2015,Pavlosiuk2016a,Pavlosiuk2016b,Hou2015b}   
This increase weakens when $B$ deviates from the transverse configuration ($\theta=90^\circ$), and is the smallest when $B$ is parallel to the electrical current (longitudinal measurement geometry, $\rm\bf{j\parallel B}$, $\theta=0^\circ$). 
In the longitudinal geometry, MR demonstrates the smallest values for both studied compounds (black solid lines in Fig.~\ref{Fig7}a,b).
After attaining the local maximum, MR decreases with increasing $B$ and even reaches the negative values. 
In $B\approx5.5$\,T for TbPtBi and $B\approx2.5$\,T for HoPtBi, inflection points occur, at which MR reaches its minimum values, $-42.6\%$ and $-23.3\%$ for TbPtBi and HoPtBi, respectively. 
In stronger magnetic fields, MR of both compounds continues to increase with increasing $B$. 
For TbPtBi it remains negative up to 14\,T, and for HoPtBi it becomes positive above 8.65\,T.
The negative MR is still observed at $\theta=30^\circ$ and at $\theta=60^\circ$ for TbPtBi and HoPtBi, respectively. 
The observation of negative LMR suggests the occurrence of CMA in both compounds.
To check if our LMR data are not affected by the parasitic current-jetting effect, we performed a squeeze test\cite{Liang2018a} (for details, see the supplementary material), which confirmed that the observed LMR is due to the intrinsic CMA.
The temperature region at which negative LMR was observed and the magnitude of the effect for both our samples are similar to those recently reported in Ref.~\onlinecite{Chen2020a}, but compared to the results in Refs.~\onlinecite{Singha2019a,Zhu2020}, we observed a much more pronounced effect for our TbPtBi sample occurring also at higher temperatures. 
When the magnetic field direction deviates from $\theta=0^{\circ}$, MR increases linearly
with increasing $\theta$ in the entire range of the magnetic field for TbPtBi, yet only in weak magnetic fields in the case of HoPtBi.
In stronger $B$, MR of the latter compound depends on $\theta$ in a more complex way; for example, in $B>5$\,T, MR measured at $\theta=60^\circ$ is larger than that recorded at $\theta=90^\circ$. 
This is clearly visible in Fig.~\ref{Fig9}, where the anisotropic magnetoresistance (AMR = $[\rho(\theta)-\rho(90^{\circ})]/\rho(90^{\circ})$) is plotted as a function of $\theta$.

To track the evolution of CMA with temperature, we performed the measurements of LMR at several different temperatures (see Fig.~\ref{Fig7}c,d). 
With increasing $T$, LMR increases, but it still demonstrates negative values, for TbPtBi up to at least $T=50$\,K, and for HoPtBi up to at least $T=10$\,K. 
Interestingly, for TbPtBi negative LMR was observed even at $T=2$\,K, which is below $T_N$. 
It may suggest that the antiferromagnetic ordering hardly affects the topologically non-trivial electronic structure of TbPtBi.      

According to the theory,~\cite{Son2013} if CMA takes place, there should occur a contribution to the longitudinal conductance that is proportional to $B^{2}$.
However, as can be inferred from Fig.~\ref{Fig7}c,d, the longitudinal electrical resistivity as a function of magnetic field, $\rho_L(B)$, varies with the magnetic field in a much more complex manner. 
At low temperatures, $\rho_L$ first rapidly increases due to WAL and then decreases due to CMA.
Eventually, $\rho_L$ increases again, probably due to the specific structure of the Fermi surface. 
There are several literature reports on topological semimetals, where LMR was considered to comprise different contributions.~\cite{Huang2015,Zhang2016} 
In order to describe quantitatively our experimental data we used the following formula:
\begin{equation}
\rho_L(B)=1/[(1+C_WB^2)\cdot\sigma_{WAL}(B)+\sigma_n(B)],
\label{CMA_eq}
\end{equation}      
where, $C_W$ stands for the chiral coefficient, $\sigma_{WAL}(B)=a\sqrt{B}+\sigma_0$ corresponds to the electrical conductivity due to the WAL effect, and $\sigma_n(B)=(\rho_0+AB^2)^{-1}$ is the electrical conductivity originating from topologically trivial electronic bands. 
As demonstrated in Fig.~\ref{Fig8}, the above formula describes well our experimental data in the entire range of magnetic fields. 
For each temperature, we obtained a negative value of the parameter $a$, indicating that the quantum correction to the electrical conductivity is due to WAL. 
We also found that the chiral coefficient gradually decreases with increasing temperature, and vanishes near $T=100$\,K. 
This reflects the weakening of CMA with increasing temperature, as observed before for quite a few other topological semimetals.~\cite{Huang2015,Lv2017,Zhang2016,Pavlosiuk2020_ScPtBi,Guo2018}
The obtained {$C_W(T)$ variation can be well approximated with the following relation:
\begin{equation}
C_W\propto\nu^3_F\tau_v/(T^2+\mu^2/\pi^2),
\label{CMA_eq2}
\end{equation}  
where, $\nu_F$ is the Fermi velocity, $\tau_v$ corresponds to the chirality-changing scattering time and $\mu$ stands for the chemical potential.~\cite{Guo2018} 
From the least-square fitting with this equation we obtained $\nu^3_F\tau_v=52\,\rm{m}^3\rm{s}^{-2}$ and $\mu=1.2$\,meV; both values are close to those reported for YbPtBi.~\cite{Guo2018}  
We attempted to apply the same model for the $\rho_L(B)$ data of HoPtBi, however the so-obtained fit was rather unsatisfactory. 
This result indicates that in the latter compound the $\sigma_n(B)$ dependence exhibits a more complex behavior.

\begin{figure}[t]
	\includegraphics[width=0.46\textwidth]{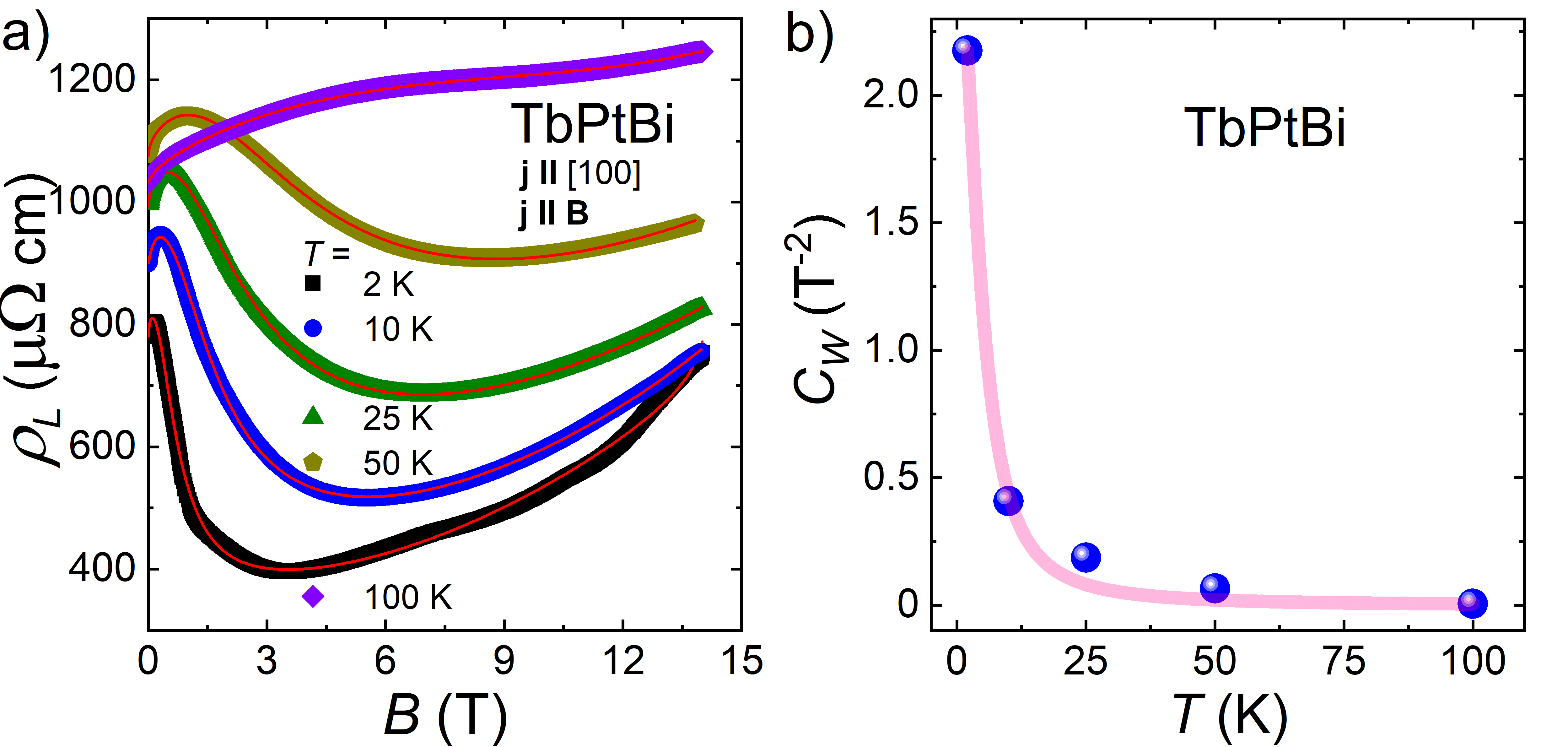}
	\caption{(a) Longitudinal electrical resistivity as a function of magnetic field measured at several different temperatures. Solid red lines are fits to Eq.~\ref{CMA_eq}. (b) The temperature dependence of the chiral coefficient. The red solid line represents the fit of Eq.~\ref{CMA_eq2} to the experimental data.  
		\label{Fig8}}
\end{figure}

According to the literature, the formation of Weyl nodes in GdPtBi is governed by the reconstruction of the electronic band structure induced by the external magnetic field via the Zeeman effect,~\cite{Hirschberger2016a} or by the exchange field due to the Gd magnetic moments.~\cite{Shekhar2018}
The latter scenario was developed based on the comparison of the electrical transport properties of antiferromagnetic GdPtBi and those of the nonmagnetic counterpart YPtBi. 
While the latter phase shows no features of CMA, our recent results showed that another nonmagnetic half-Heusler phase ScPtBi also demonstrates negative LMR, planar Hall effect and angular narrowing of negative LMR, and all these effects are caused by CMA.~\cite{Pavlosiuk2020_ScPtBi}
Therefore, we cannot exclude that in TbPtBi and HoPtBi, the Weyl nodes appear due to the Zeeman interaction. 
	
Strong dependence of MR on the angle $\theta$ implies considerable AMR. 
Fig.~\ref{Fig9}a displays AMR($\theta$) dependences, taken in $B=14$\,T at several temperatures from the range 2-300\,K. 
For $T\ge 200$\,K, AMR($\theta$) can be well described using the standard formula $\rm{AMR}\propto\cos^2$$(\theta)$ (see the supplementary material); however, at lower temperatures this law was no longer followed. 
At temperatures up to at least 10\,K, local minima at $\theta=90^\circ$ and $\theta=270^\circ$ form between the pairs of local maxima at $\theta=60^\circ$, $\theta=120^\circ$ and $\theta=240^\circ$, $\theta=300^\circ$, respectively. 
Moreover, clear kinks occur (e.g., near $\theta=35^\circ,145^\circ,215^\circ$ and $325^\circ$ at $T=2$\,K), whose positions change with increasing temperature. The latter features entirely disappear at $T=100$\,K. 
A similar behavior of AMR was reported before for GdPtBi~\cite{Kumar2018} and DyPdBi.~\cite{Pavlosiuk2019a} 
In our recent work,~\cite{Pavlosiuk2019a} we attributed this effect to the crystalline anisotropy of the single crystals examined, and we suppose that such an explanation is suitable also for HoPtBi.
However, it should be recalled that a similar angle-dependent magnetoresistance was reported also for the archetypal nodal-line Dirac semimetal ZrSiS, and the authors called such an unusual phenomenon butterfly magnetoresistance.~\cite{Ali2016}
It was subsequently shown~\cite{Novak2019a,Voerman2019} that this unusual MR can be understood in the scope of the Fermi surface architecture, namely, the formation of open orbits and existence of hole- and electron-like carriers, subjected to the Zeeman effect.
Remarkably, the MR($\theta$) data of HoPtBi, presented in a polar coordinate system (see Fig.~\ref{Fig9}b), have below 50 K a butterfly-like shape, very similar to those reported for ZrSiS in Ref.~\onlinecite{Ali2016}.  
	
In $B=14$\,T, HoPtBi exhibits the giant values of AMR in the entire temperature interval covered. 
For example, AMR = $-55\%$ at $T=300$\,K. 
Large AMR is a property that can be used in practical applications, if the effect occurs near room temperature and in small magnetic field. 
To estimate the possible application potential of HoPtBi we investigated its AMR at $T=300$\,K in several values of the applied magnetic field. 
As apparent from Fig.~\ref{Fig9}c,d, with a decrease in magnetic field, the magnitude of AMR systematically decreases. 
Although AMR is still observed in a relatively small magnetic field of 0.1\,T, its value is only $-1.8\!\times\!10^{-2}\%$, i.e., notably smaller than the AMR of permalloy, a material that is commonly used in AMR-based devices.~\cite{Grosz2017}  

\begin{figure}[t]
	\includegraphics[width=0.46\textwidth]{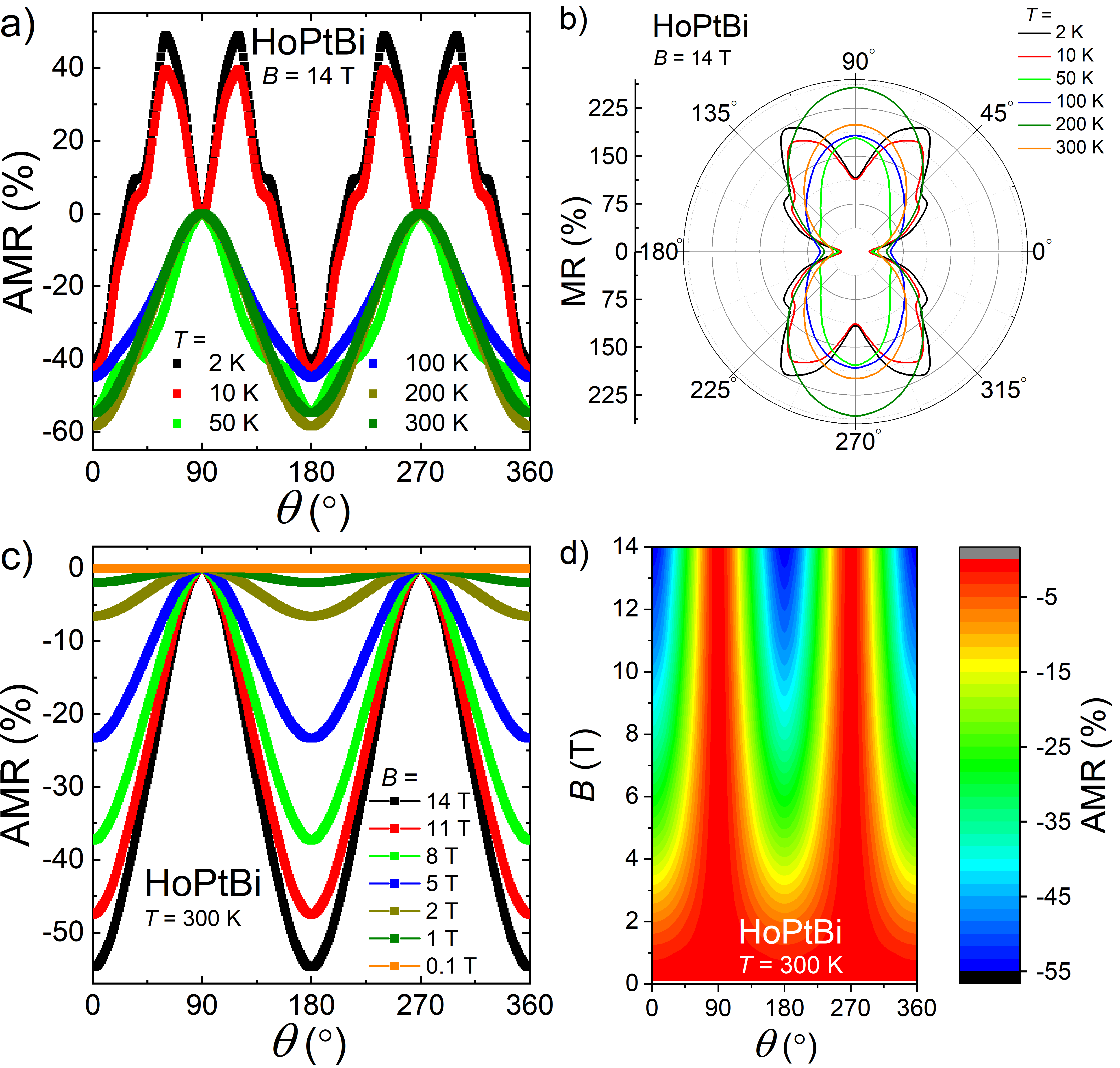}
	\caption{(a) Angular dependence of the anisotropic magnetoresistance of HoPtBi measured in $B=14$\,T at several different temperatures. (b) The polar plot of the magnetoresistance of HoPtBi at several temperatures. (c) The angular dependence of the anisotropic magnetoresistance at $T=300$\,K measured in different applied magnetic field. (d) The two-dimensional map representation of the AMR data shown in panel (c). 
		\label{Fig9}}
\end{figure}

\section{Conclusions}
	
This paper presents the magnetotransport properties, which can be attributed to the possible existence of topologically non-trivial electronic states in half-Heusler compounds HoPtBi and TbPtBi.
We also confirmed the existence of antiferromagnetic ordering below N\'{e}el temperatures $T_N= 3.36$\,K and 1.26\,K for TbPtBi and HoPtBi, respectively. 
Both compounds exhibit considerable values of the transverse magnetoresistance, that disobeys Kohler's rule. 
Analysis of the SdH effect, observed for HoPtBi, indicates the existence of two principal frequencies and relatively small effective masses of carriers.   
Negative LMR was observed for both studied compounds most likely due to CMA, suggesting that the investigated materials can be topological semimetals. 
Furthermore, TbPtBi and HoPtBi were found to exhibit the AHE. 
In TbPtBi, both effects were observed at temperatures below and above $T_N$, excluding the antiferromagnetic order as a possible source of negative LMR and AHE.
While the electrical transport properties of HoPtBi and TbPtBi comply with the possible existence of massless Weyl fermions, their origin remains unclear and further theoretical and experimental work is necessary to clarify the issue.
\\
\\{\bf{SUPPLEMENTARY MATERIAL}}\\
\\
See the supplementary materials for (i) Laue patterns; (ii) Kohler scaling of magnetoresistance; (iii) details of Hall conductivity analysis; and (iv) angle dependent electrical resistivity of HoPtBi at $T=200$\,K and $T=300$\,K.\\ 

\begin{acknowledgments}
We thank E. Bukowska and D. Szyma\'{n}ski for performing powder X-ray diffractography and EDS analysis, respectively. This research was funded by the National Science Centre of Poland (Grant No. 2015/18/A/ST3/00057). O.P. was supported by the Foundation for Polish Science (FNP), program START 66.2020.
\end{acknowledgments}

\nocite{*}

\newpage
\noindent\begin{large}{\bf Supplementary Material} \end{large}
%
\setcounter{figure}{0}
\setcounter{table}{0}
\setcounter{equation}{0}
\renewcommand{\thefigure}{S\arabic{figure}}
\renewcommand{\thetable}{S\arabic{table}}
\renewcommand{\theequation}{S\arabic{equation}}
\subsection*{Laue diffraction}
\begin{figure}[h]
	\includegraphics[width=0.235\textwidth]{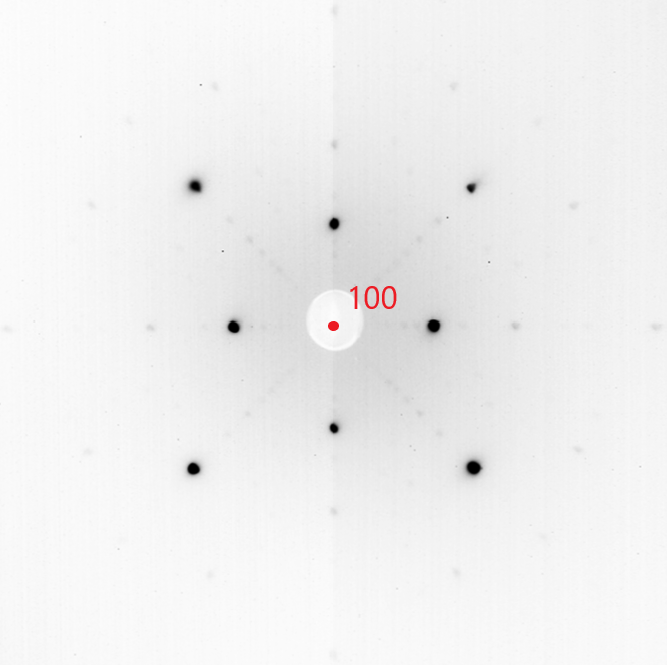}	
	\includegraphics[width=0.235\textwidth]{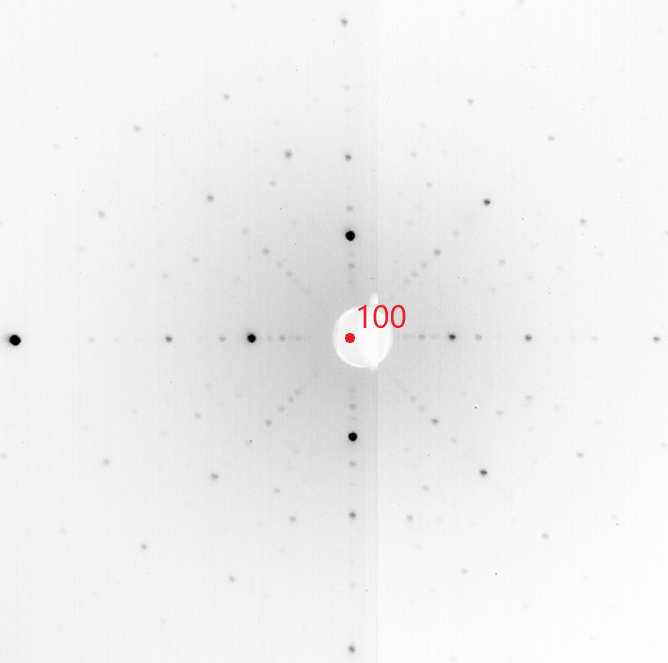}
	\caption{Backscattering Laue patterns of TbPtBi (a) and HoPtBi (b) for X-ray incident beam directed along the [100] crystallographic direction.}
	\label{Laue_img}
\end{figure}
\subsection*{Kohler scaling of magnetoresistance}
\begin{figure}[h]
	\includegraphics[width=0.4\textwidth]{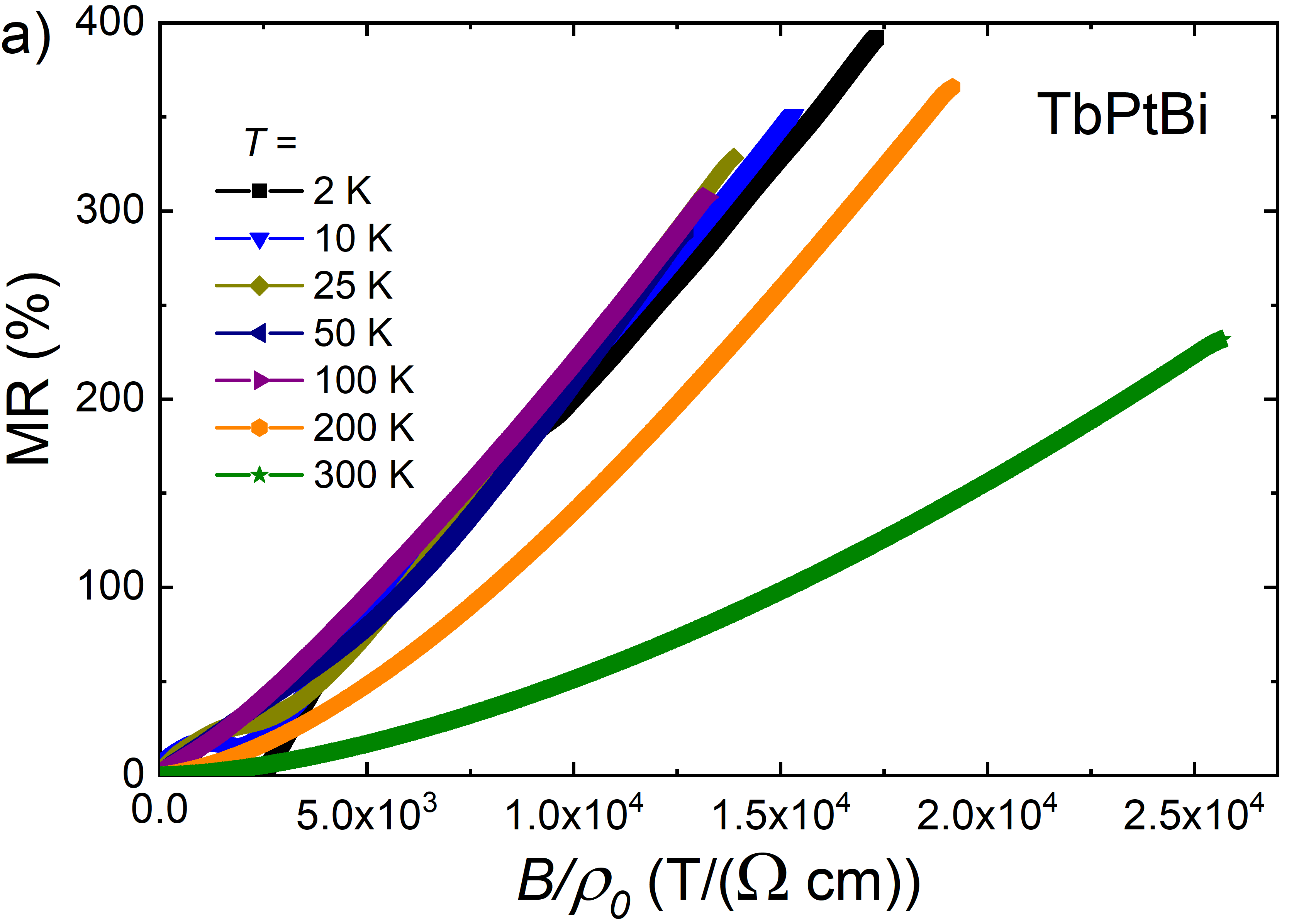}
	\includegraphics[width=0.4\textwidth]{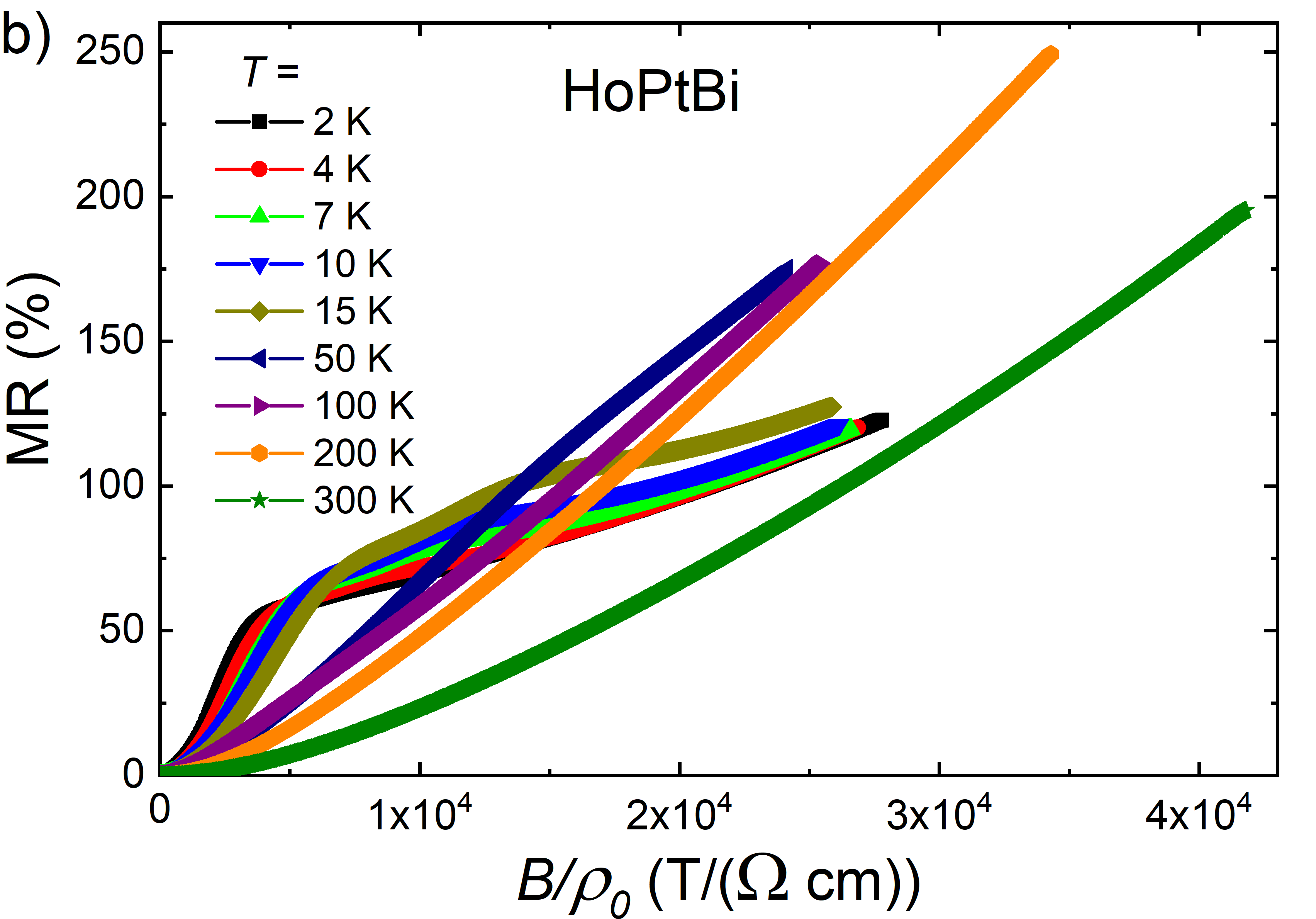}
	\caption{Kohler's plot of the magnetoresistance data for TbPtBi (a) and HoPtBi (b). 
		\label{Kohler}}
\end{figure}
According to the semiclassical transport theory, in a metal with single carrier type and single scattering time, magnetoresistance should follow the Kohler's rule: ${\rm MR}\!=\!f(B/\rho(0))$, where $\rho(0)$ stands for the electrical resistivity in zero magnetic field, and $f$ is the function determined only by geometry of experiment and nature of metal.\cite{Ziman2001_S}
Assuming that $\rho_0=1/(ne\mu)$, where, $e$ is the elementary charge, $n$ stands for the carrier concentration, and $\mu$ is the carrier mobility, it is obvious that if $n$ and $\mu$ are temperature independent the Kohler's law is obeyed, MR($B$) isotherms or MR($T$) curves collapse onto a single curve. 
However, as it was shown by Wang et al.,\cite{Wang2015d_S} this rule can be extended to semimetals as well, but only to those that demonstrate perfect carrier compensation. 
In the case of some difference between the concentrations of carriers of both types, the rule can be fulfilled if only both mobilities are small or at least one of them is small. 
Besides, carrier concentrations and the relation between both mobilities ($\mu_e/\mu_h$) should be temperature independent.

The Kohler's plots constructed for TbPtBi and HoPtBi are shown in Fig.~\ref{Kohler}. 
In the whole range of studied temperatures, there are no two curves, among all ${\rm MR}(B/\rho_0)$ dependences, that completely collapse onto each other.
This means that Kohler's rule is not fulfilled for both studied compounds. 
\subsection*{Hall conductivity analysis}
\begin{figure}[h]
	\includegraphics[width=0.4\textwidth]{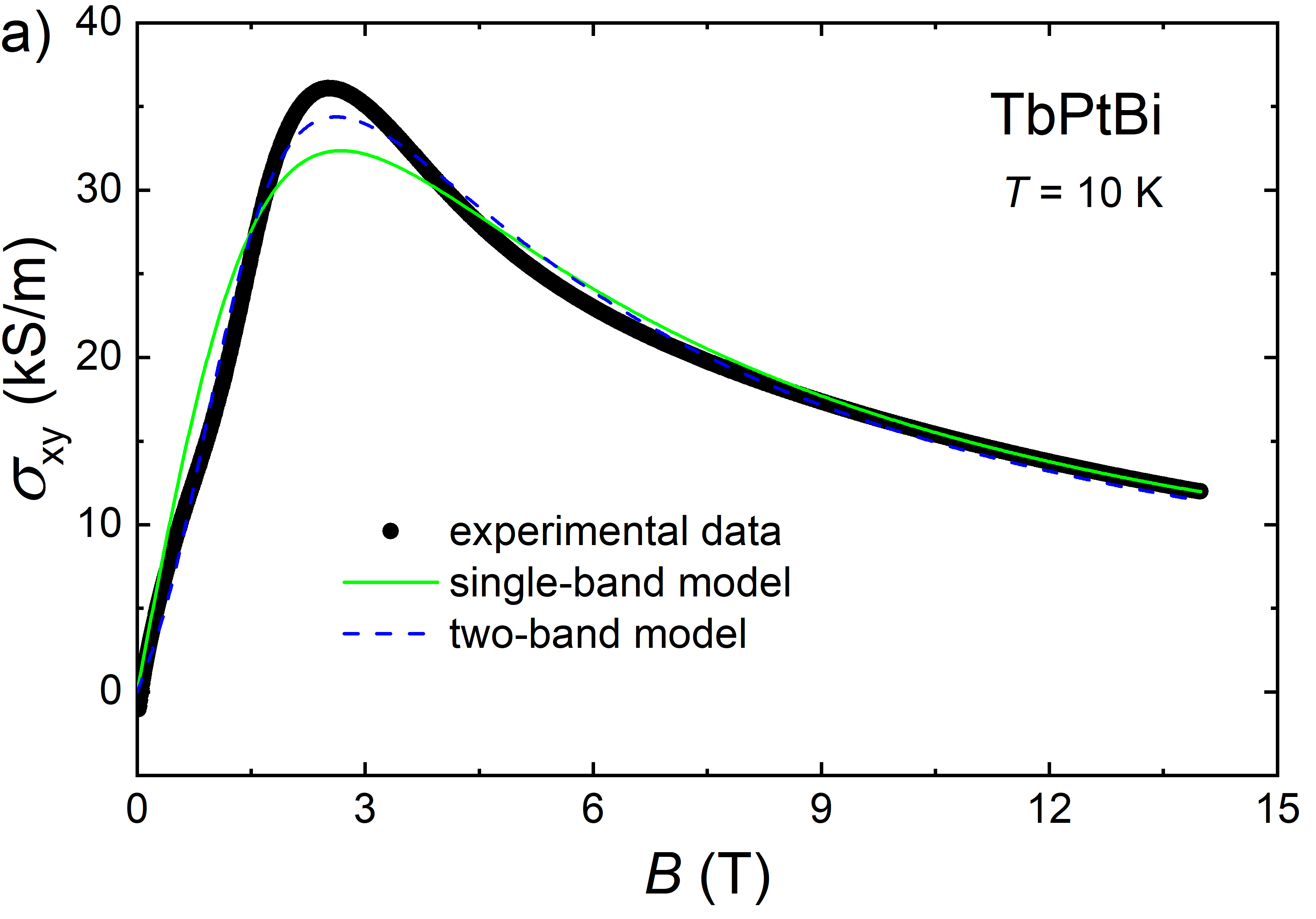}
	\includegraphics[width=0.4\textwidth]{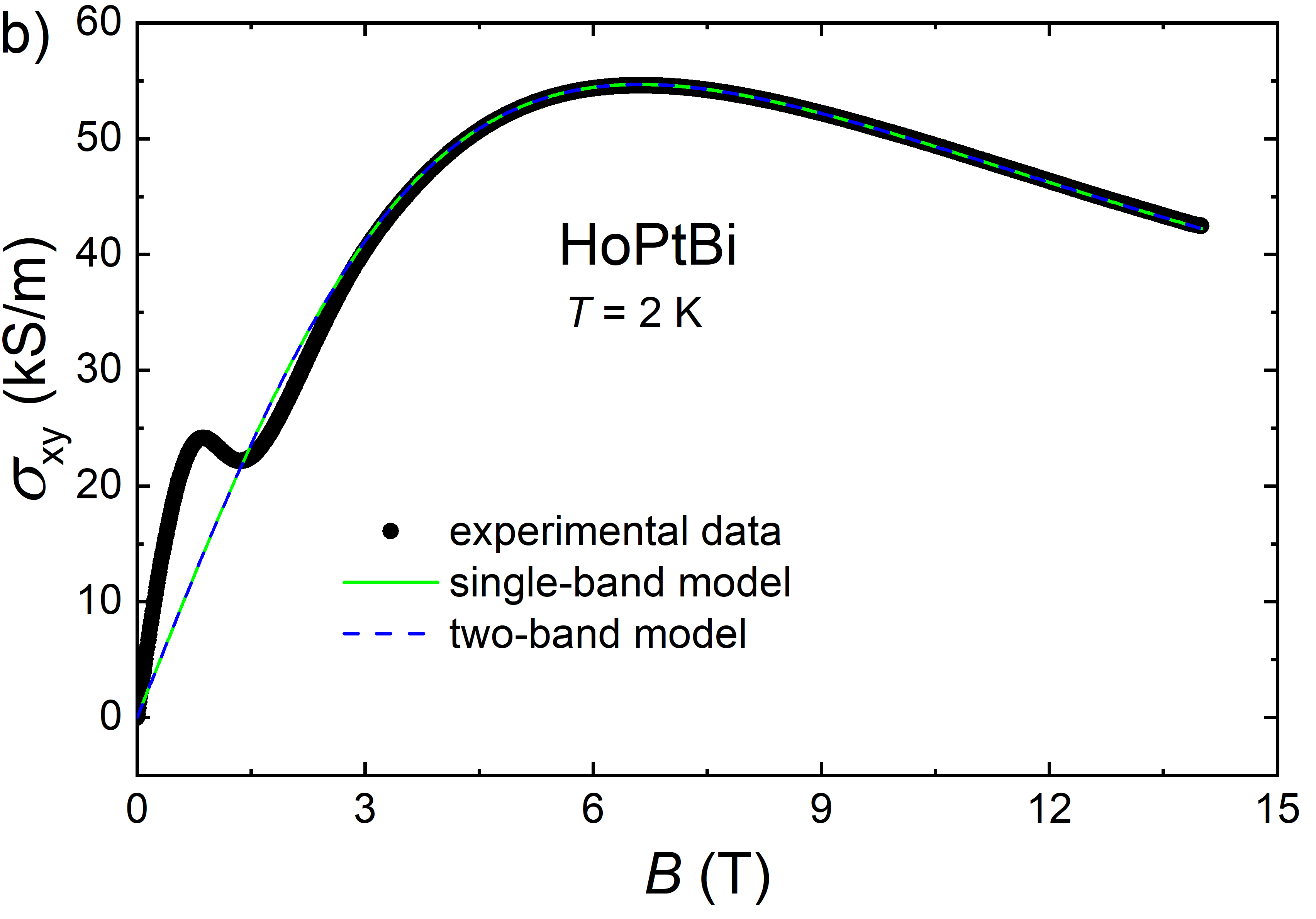}
	\caption{Hall conductivity as a function of magnetic field at $T=10$\,K for TbPtBi (a) and at $T=2$\,K for HoPtBi (b). Blue dashed and green solid lines correspond to the results of fitting two- and single-band models to the experimental data.}
	\label{Hall_img_1}
\end{figure}

We tried to describe the results of Hall effect measurements using the standard single- and two-band Drude models. 
The results of fitting of Eq.~3 (see main text) to the experimental data are presented in Fig.~\ref{Hall_img_1}a,b. 
The analysis was performed on data taken at $T=10$\,K and at $T=2$\,K for TbPtBi and HoPtBi, respectively, these $T$ are above corresponding $T_N$.  
None of the models can in a proper way describe the experimental results, besides, two-band model is over-parametrised relative to the data set. 
This indicates the occurrence of some additional contribution to the Hall effect data, which can be attributed to the anomalous Hall effect or changes of electronic band structure induced by magnetic field, as in GdPtBi.\cite{Hirschberger2016a_S,Shekhar2018_S}

\begin{figure}[b]
	\includegraphics[width=0.23\textwidth]{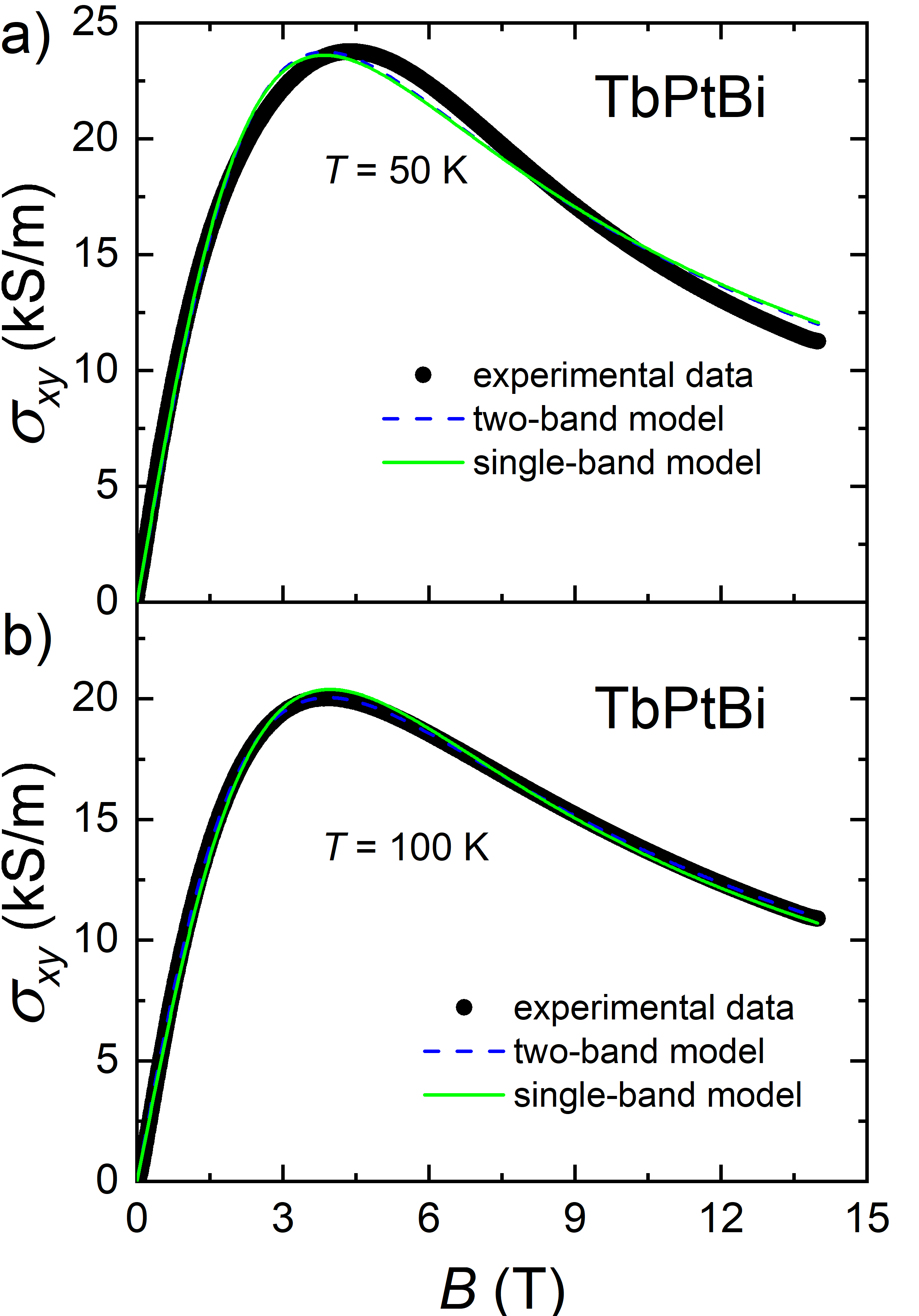}
	\includegraphics[width=0.23\textwidth]{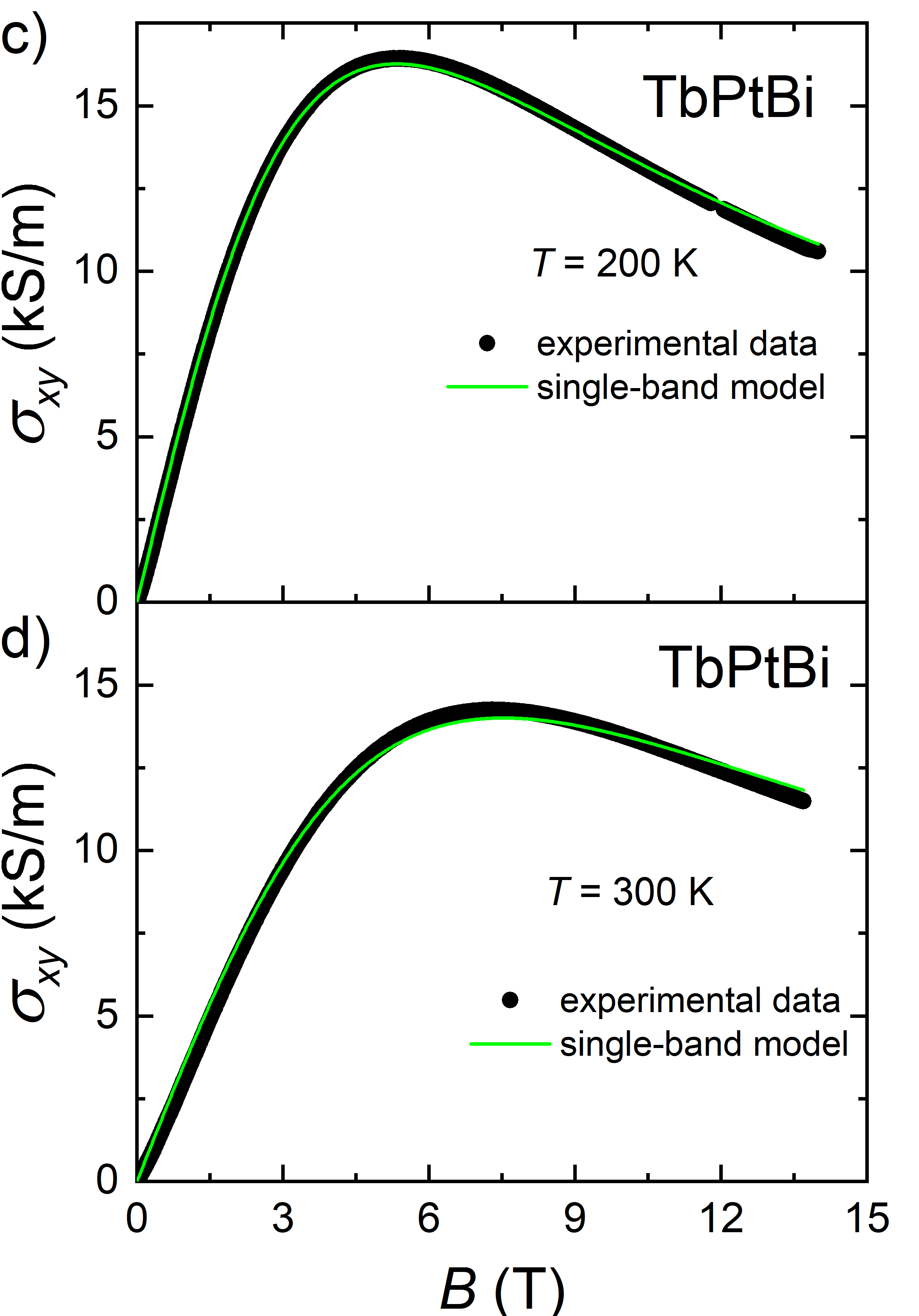}
	\includegraphics[width=0.23\textwidth]{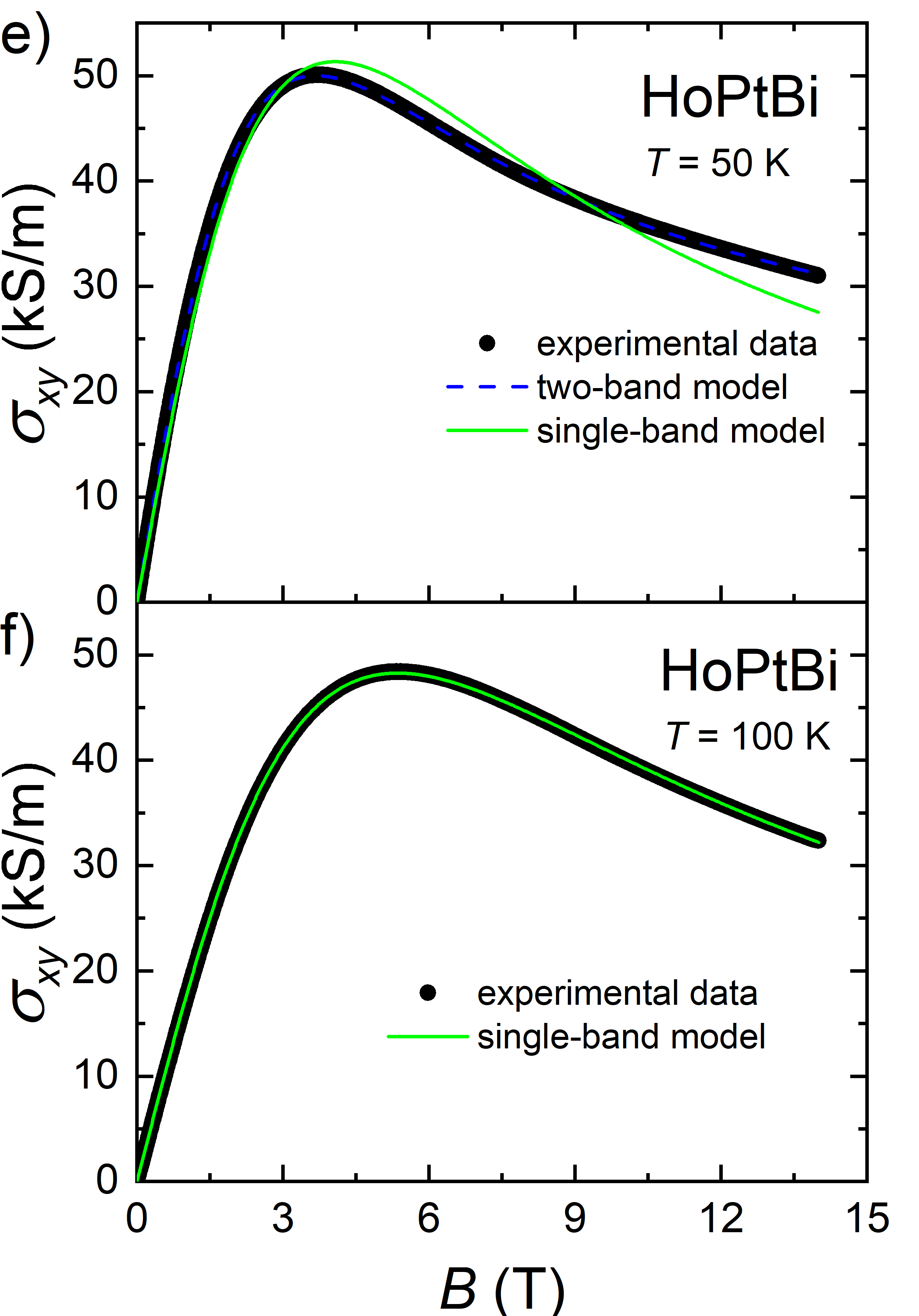}
	\includegraphics[width=0.23\textwidth]{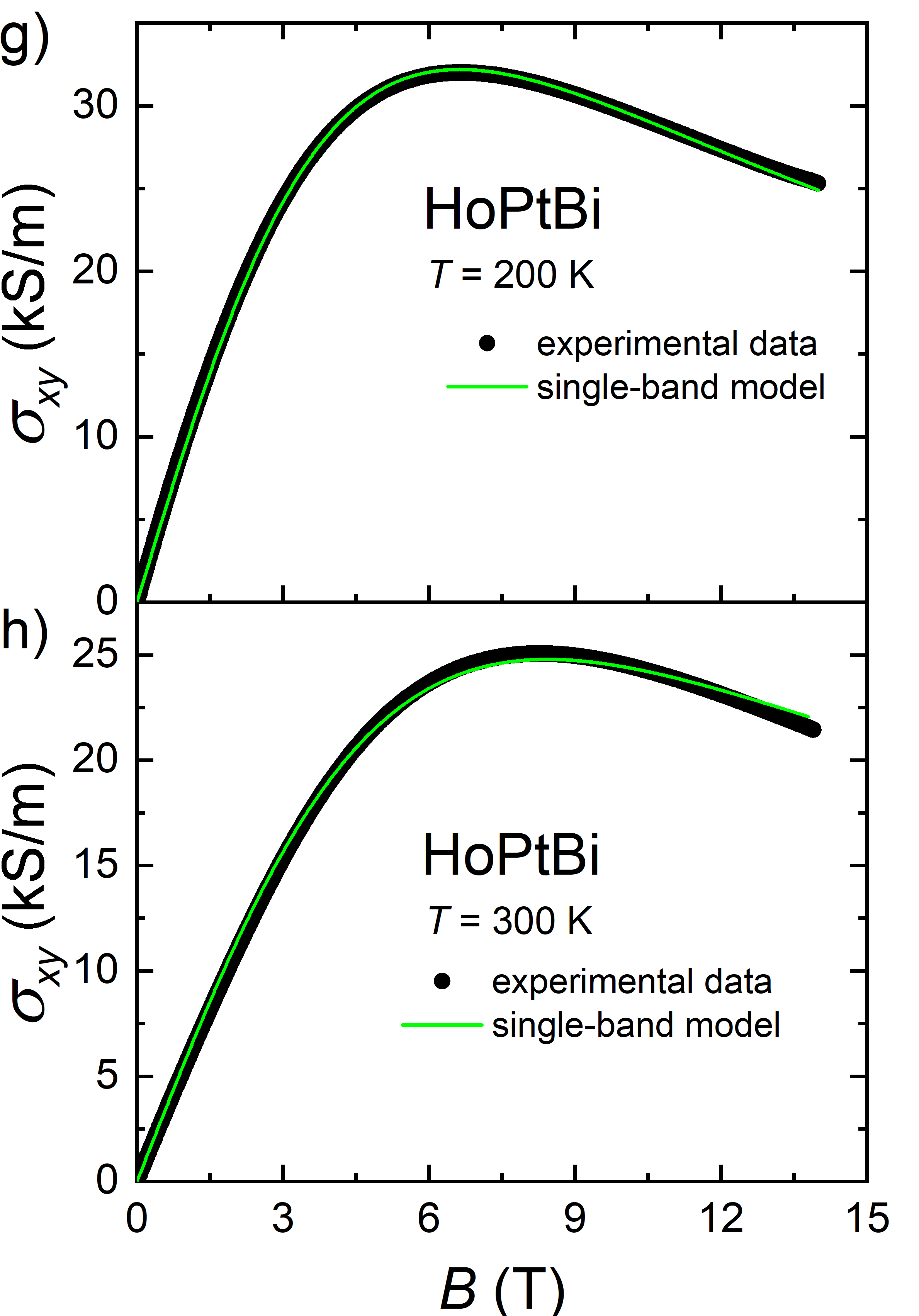}
	\caption{Hall conductivity versus magnetic field measured at several temperatures for TbPtBi (a-d) and HoPtBi (e-h). Green solid and blue dashed lines are the fits of the single- and two-band models.}
	\label{Hall_img_2}
\end{figure}

Fig.~\ref{Hall_img_2} shows the results of fitting the Drude models to the electrical conductivity data obtained for TbPtBi and HoPtBi at several constant temperatures from the range $50-300$\,K. 
In the temperature range $100-300$\,K, $\sigma_{xy}(B)$ dependences can be approximated with the single-band Drude model, green solid lines in Fig.~\ref{Hall_img_2}. 
When we tried to use two-band model to improve the quality of fits, we found that the model is over-parametrised relative to the experimental data. 
The only exception is $\sigma_{xy}(B)$ of TbPtBi taken at 100\,K, the dependence can be fitted with two-band model.
$\sigma_{xy}(B)$ of TbPtBi obtained at $T=50$\,K cannot be properly described by any of the above models (see Fig.~\ref{Hall_img_2}a), probably due to the considerable contribution of the anomalous Hall effect at this temperature.   
In the case of HoPtBi, its $\sigma_{xy}(B)$ at $T=50$\,K can be well approximated by two-band model (see Fig.~\ref{Hall_img_2}e). 
All obtained fitting parameters are gathered in Table~\ref{Hall_parameters_TbPtBi} and Table~\ref{Hall_parameters_HoPtBi}.
Apparently, both carrier concentration and carrier mobility are considerably dependent on temperature.  

\begin{table}[h]
	\centering
	\caption{Parameters obtained for TbPtBi from single- and two-band Drude models fitting to the Hall conductivity shown in Fig.~\ref{Hall_img_2}; $n_{h1}$ and $n_{h2}$ - hole concentrations, $\mu_{h1}$ and $\mu_{h2}$ - hole mobilities.}
	\begin{tabular*}{0.49\textwidth}{@{\extracolsep{\fill}}*{5}{c}} \hline\hline
		$T$ & $n_{h1}$ & $\mu_{h1}$& $n_{h2}$ & $\mu_{h2}$ \\
		(K) & (cm$^{-3}$) & ($\rm{cm^2V^{-1}s^{-1}}$) & (cm$^{-3}$) & ($\rm{cm^2V^{-1}s^{-1}}$) \\\hline
		100 & $3.19\!\times\!10^{17}$ & $3491$ & $7.3\!\times\!10^{17}$ & $2028$\\
		100 & $1.01\!\times\!10^{18}$ & $2518$ & - & -\\
		200 & $1.08\!\times\!10^{18}$ & $1875$ & - & -\\
		300 & $1.32\!\times\!10^{18}$ & $1327$ & - & -\\
		\hline\hline
	\end{tabular*}\label{Hall_parameters_TbPtBi}
\end{table}

\begin{table}[h]
	\centering
	\caption{Parameters obtained for HoPtBi from single- and two-bands Drude models fitting to the Hall conductivity shown in Fig.~\ref{Hall_img_2}; $n_{h1}$ and $n_{h2}$ - hole concentrations, $\mu_{h1}$ and $\mu_{h2}$ - hole mobilities.}
	\begin{tabular*}{0.49\textwidth}{@{\extracolsep{\fill}}*{5}{c}} \hline\hline
		$T$ & $n_{h1}$ & $\mu_{h1}$& $n_{h2}$ & $\mu_{h2}$ \\
		(K) & (cm$^{-3}$) & ($\rm{cm^2V^{-1}s^{-1}}$) & (cm$^{-3}$) & ($\rm{cm^2V^{-1}s^{-1}}$) \\\hline
		50 & $2.01\!\times\!10^{18}$ & $2909$ & $3.43\!\times\!10^{18}$ & $404$ \\
		100 & $3.23\!\times\!10^{18}$ & $1866$ & - & - \\
		200 & $2.67\!\times\!10^{18}$ & $1508$ & - & - \\
		300 & $2.61\!\times\!10^{18}$ & $1184$ & - & - \\
		\hline\hline
	\end{tabular*}\label{Hall_parameters_HoPtBi}
\end{table}

\subsection*{Squeeze test}

The concept of the squeeze test, which allows to verify if the measurements of longitudinal magnetoresistance are burdened with current-jetting effect was introduced in Ref.~\onlinecite{Liang2018a}. 
Following that paper we performed such test on our TbPtBi sample. 
In the ideal squeeze test, all electrical contacts (one pair of current contacts and two pairs of voltage contacts) should be point-like, as it is schematically shown in the lower inset to Fig.~\ref{squeeze_Fig}a. 
The upper inset is the photograph of the sample, with attached electrical contacts, that was tested. 
We also carried out the squeeze test on the same sample with the same configuration of two pairs of voltage leads, but with the stripe-like current contacts, they were attached along the whole width of the sample (see insets to Fig.~\ref{squeeze_Fig}b). 
The aim of this attachment of current leads was to reduce the current-jetting effect (in such configuration the current distribution should be more homogeneous).  
For both measurement geometries, the obtained data are rather similar (see main panel of Fig.~\ref{squeeze_Fig}a and Fig.~\ref{squeeze_Fig}b) and support our conclusion that observed longitudinal magnetoresistance is dominated by the chiral magnetic anomaly, and not the current-jetting effect.
In both cases, we obtained longitudinal magnetoresistances, measured in spine and edge configurations, that demonstrate negative values in some ranges of magnetic field. 
This behaviour is in analogy to the results reported for Na$_3$Bi and GdPtBi in Ref.~\onlinecite{Liang2018a}. 
The negative longitudinal magnetoresistance of both these compounds originate from chiral magnetic anomaly, but not form current-jetting effect.

\begin{figure}[h]
	\includegraphics[width=0.4\textwidth]{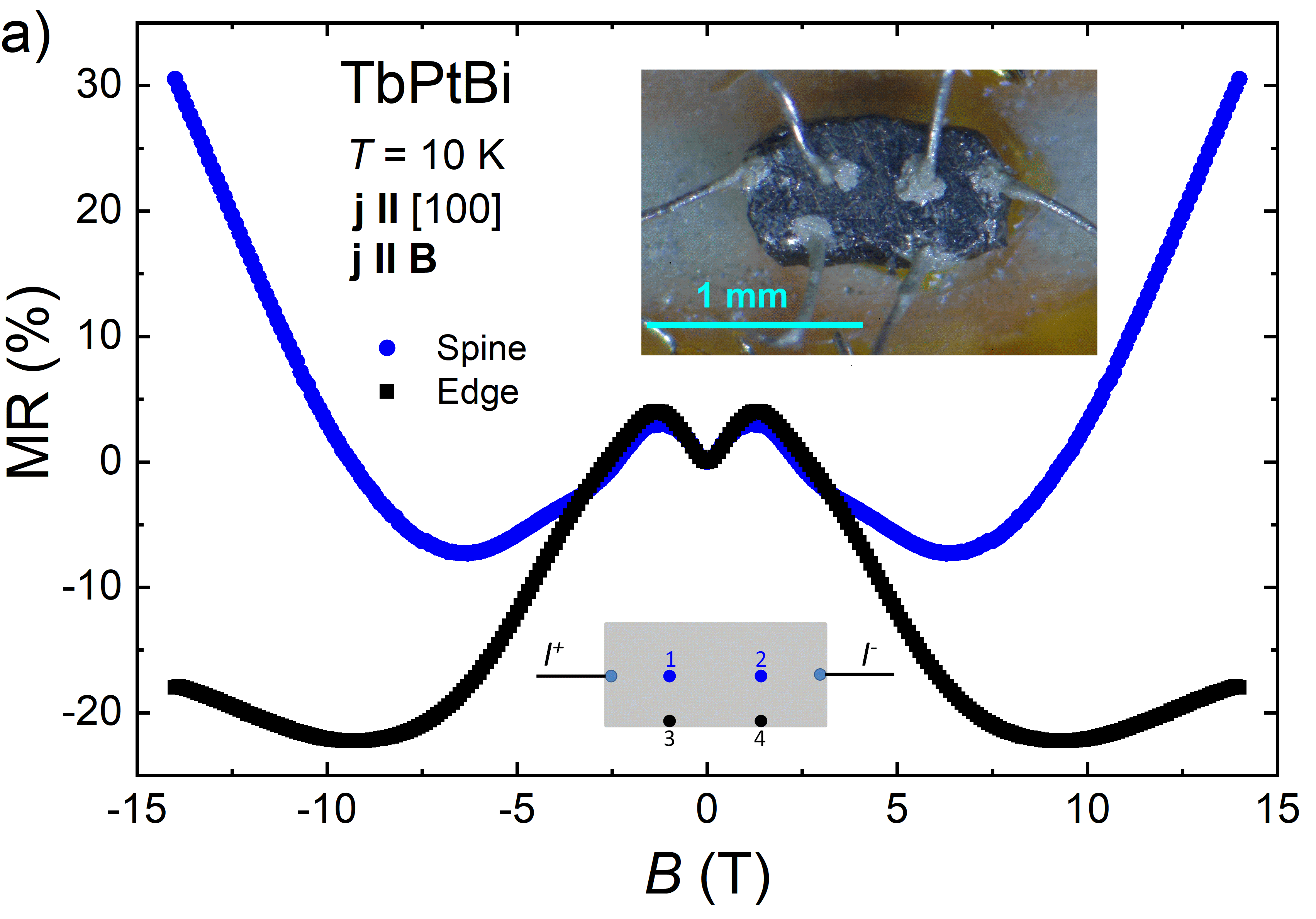}
	\includegraphics[width=0.4\textwidth]{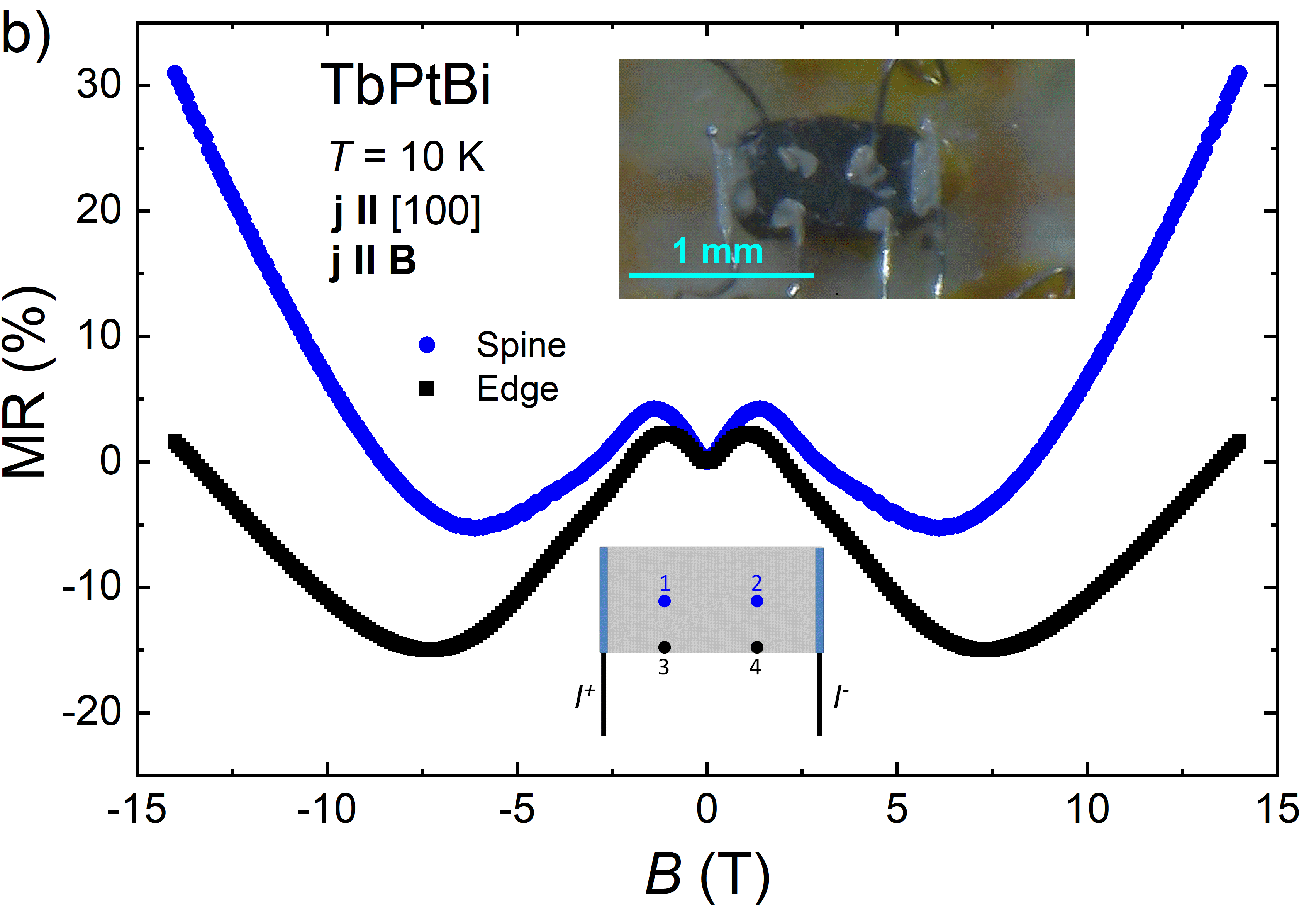}
	\caption{Results of squeeze test performed at $T=10$\,K in TbPtBi. 
		Panels (a) and (b) show the magnetic field dependence of longitudinal ($\rm\bf{j\!\parallel\!B}$) magnetoresistance obtained for samples with point-like and stripe-like current contacts, respectively. Upper insets display photographs of measured samples and lower insets show schematically the measurement geometries. Two pairs of voltage contacts were used, 1-2 pair located in the middle of the sample on the same line as current contacts (spine configuration) and 3-4 pair placed on the edge of the sample (edge configuration).}
	\label{squeeze_Fig}
\end{figure}

\subsection*{Angle dependent electrical resistivity}

Despite that the formula (Eq.~\ref{AMR_eq}) for describing angle dependent electrical resistivity ($\rho(\theta)$) was developed for polycrystalline materials,\cite{Jan1957_S} it can also be applicable for single-crystalline substances.
\begin{equation}
\rho(\theta)=\rho_{\perp}-\Delta\rho\cos^2(\theta).
\label{AMR_eq}
\end{equation}
In the above equation, $\rho_{\perp}$ stands for the electrical resistance recorded in the transverse measurement configuration ($\rm\bf{j\!\perp\!B}$, $\theta=90^\circ$), and $\Delta\rho=\rho_{\perp}-\rho_{L}$, where $\rho_L$ is the electrical resistivity measured in the longitudinal measurement geometry ($\rm\bf{j\!\parallel\!B}$, $\theta=0^\circ$).
We found that this formula well describes the obtained data at $T\geq200$\,K. 
The results of fitting Eq.~\ref{AMR_eq} to the experimental data are shown as red solid lines in panels (a), (b) and (c) of Fig.~\ref{AMR}. 
We found that at $T=300$\,K, magnetic field dependence of $\Delta\rho$ can be well approximated by the exponential growth equation $\Delta\rho=AB^n$ with $A=0.011\,\rm{m\Omega\,cm}$ and $n=1.52$ (green line in Fig.~\ref{AMR}d). 

\begin{figure}[h]
	\includegraphics[width=0.4\textwidth]{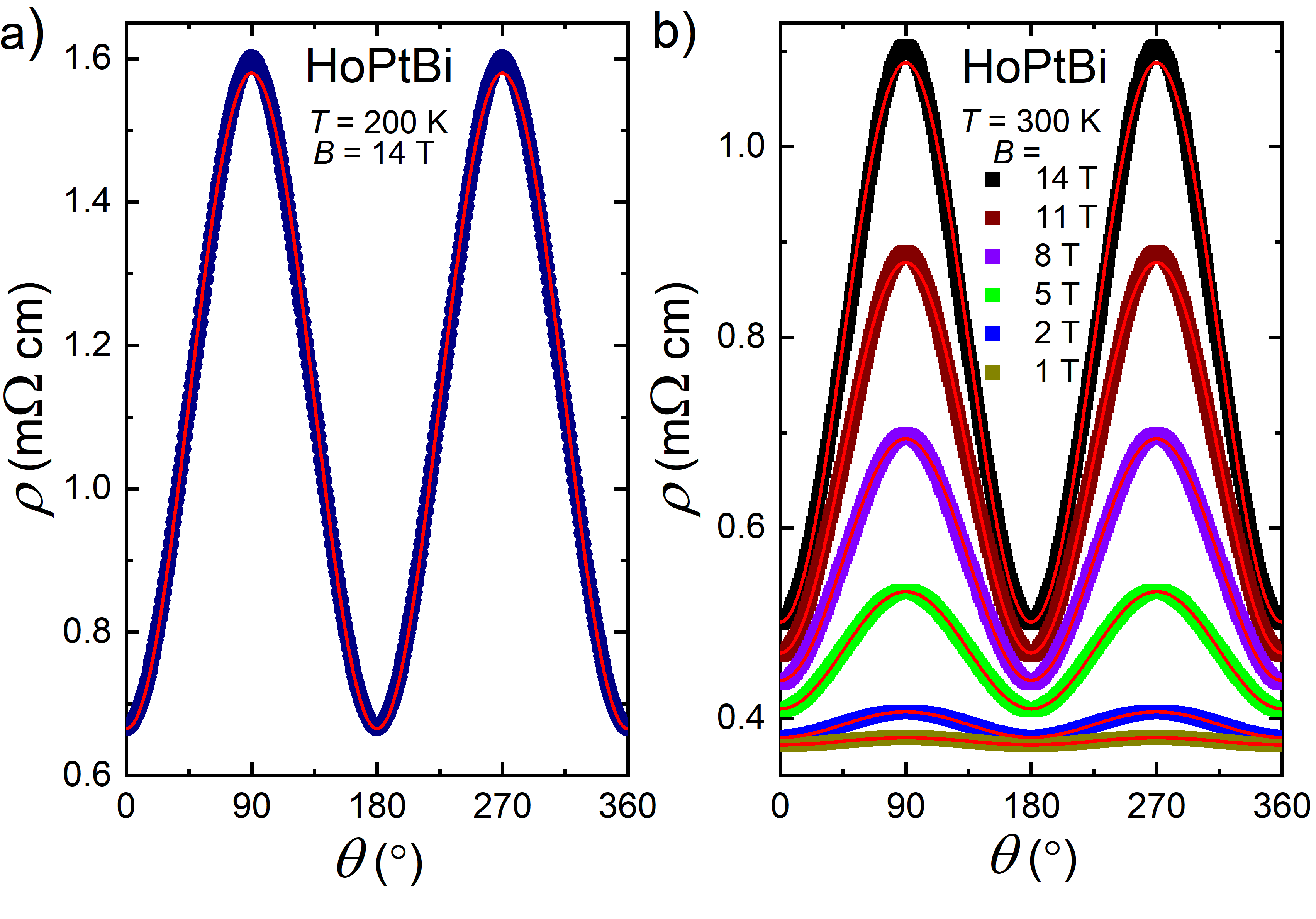}
	\includegraphics[width=0.4\textwidth]{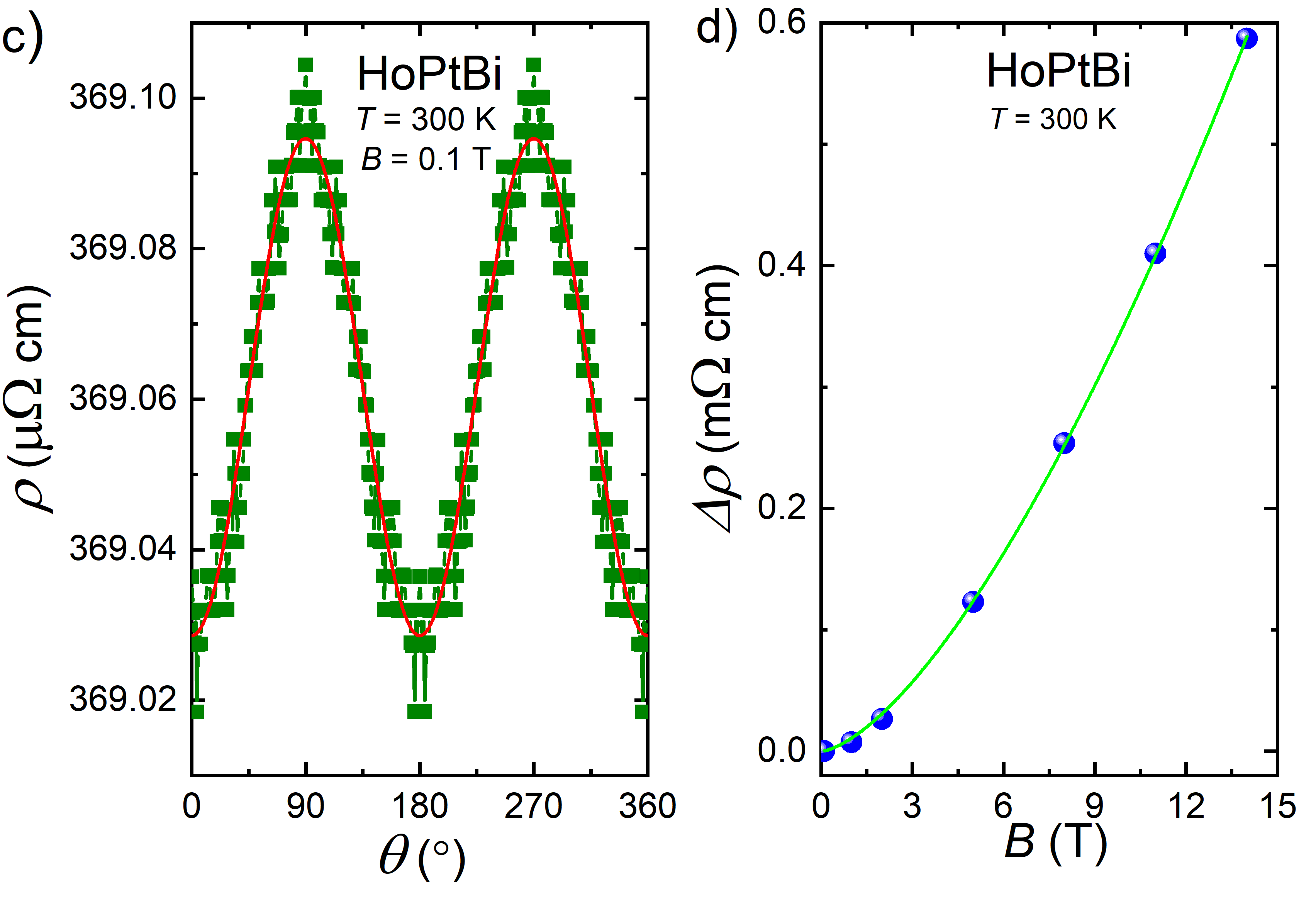}
	\caption{Angular dependence of the electrical resistivity of HoPtBi taken at $T=200$\,K in $B=14$\,T (a); at $T=300$\,K in several different values of magnetic field (b); at $T=300$\,K in $B=0.1$\,T (c). Red solid lines represent the fits with Eq.~\ref{AMR_eq}. (d) Magnetic field dependence of the resistivity anisotropy at $T=300$\,K obtained from the fits shown in panels (b) and (c). Green solid line corresponds to the fit described by expression $\Delta\rho=AB^n$.}
	\label{AMR}
\end{figure}

\vspace{0.5cm}
\noindent{\bf Supplemental references:}

\end{document}